\documentclass[acmlarge]{acmart}
\usepackage{array}
\usepackage{rotating}
\usepackage{longtable}
\usepackage{multirow}
\usepackage{lscape}
\usepackage{csquotes}
\usepackage{fontawesome}
\usepackage{ragged2e}
\usepackage{graphicx}
\usepackage{makecell}
\usepackage{booktabs}
\usepackage{caption}
\usepackage{subcaption}
\usepackage{pifont}
\newcommand{\cmark}{\ding{51}}
\newcommand{\xmark}{\ding{53}}
\usepackage{soul}
 \usepackage[edges]{forest}
 \tikzset{every shadow/.style={shadow xshift=5pt,shadow yshift=-5pt}}
\usetikzlibrary{shadows}
\AtBeginDocument{%
  \providecommand\BibTeX{{%
    \normalfont B\kern-0.5em{\scshape i\kern-0.25em b}\kern-0.8em\TeX}}}

\acmJournal{CSUR}
\acmVolume{}
\acmNumber{}
\acmArticle{}
\acmMonth{3}

\begin{document}

%%
%% The "title" command has an optional parameter,
%% allowing the author to define a "short title" to be used in page headers.
\title{Blockchain-based Digital Twins: Research Trends, Issues, and Future Challenges}

\author{Sabah Suhail}
\affiliation{%
  \institution{University of Tartu}
  \city{Tartu}
  \country{Estonia}}
\email{sabah.suhail@ut.ee}

\author{Rasheed Hussain}
\affiliation{%
  \institution{Innopolis University}
  \city{Innopolis}
  \country{Russia}}
\email{r.hussain@innopolis.ru}

\author{Raja Jurdak}
\affiliation{%
  \institution{Queensland University of Technology}
  \city{Brisbane}
  \country{Australia}}
\email{r.jurdak@qut.edu.au}

\author{Alma Oracevic}
\affiliation{%
  \institution{Innopolis University}
  \city{Innopolis}
  \country{Russia}}
\email{a.oracevic@innopolis.ru}

\author{Khaled Salah}
\affiliation{%
 \institution{Khalifa University}
 \city{Abu Dhabi}
 \country{UAE}}
\email{khaled.salah@ku.ac.ae}

\author{Raimundas Matulevičius}
\affiliation{%
  \institution{University of Tartu}
  \city{Tartu}
  \country{Estonia}}
\email{raimundas.matulevicius@ut.ee}

\author{Choong Seon Hong}
\affiliation{%
  \institution{Kyung Hee University}
  \city{Yongin}
  \country{South Korea}}
\email{cshong@khu.ac.kr}

\begin{abstract}
% Word Limit: 200
Industrial processes rely on sensory data for decision-making processes, risk assessment, and performance evaluation. Extracting actionable insights from the collected data calls for an infrastructure that can ensure the dissemination of trustworthy data. For the physical data to be trustworthy, it needs to be cross-validated through multiple sensor sources with overlapping fields of view. Cross-validated data can then be stored on the blockchain, to maintain its integrity and trustworthiness. Once trustworthy data is recorded on the blockchain, product lifecycle events can be fed into data-driven systems for process monitoring, diagnostics, and optimized control. In this regard, Digital Twins (DTs) can be leveraged
to draw intelligent conclusions from data by identifying the faults and recommending precautionary measures ahead of critical events. Empowering DTs with blockchain in industrial use-cases targets key challenges of disparate data repositories, untrustworthy data dissemination, and the need for predictive maintenance. In this survey, while highlighting the key benefits of using blockchain-based DTs, we present a comprehensive review of the state-of-the-art research results for blockchain-based DTs. Based on the current research trends, we discuss a trustworthy blockchain-based DTs framework. We highlight the role of Artificial Intelligence (AI) in blockchain-based DTs. Furthermore, we discuss current and future research and deployment challenges of blockchain-supported DTs that require further investigation.
\end{abstract}

%%
%% The code below is generated by the tool at http://dl.acm.org/ccs.cfm.
%% Please copy and paste the code instead of the example below.
%%



%%
%% Keywords. The author(s) should pick words that accurately describe
%% the work being presented. Separate the keywords with commas.

\maketitle

\section{Introduction}\label{introduction}
Traditional Industrial Control Systems (ICS) include operational technology (OT) such as supervisory control and data acquisition (SCADA) systems, programmable logic controllers (PLCs), and other field devices that are able to command and control physical processes within industrial environments~\cite{dietz2020unleashing}. In the course of Industry 4.0, this industrial infrastructure is increasingly integrated with general-purpose Information Technology (IT) systems by wireless connections with cost-effective miniaturized devices having sensing, computation, and communication capabilities to collect, process, analyze, and interpret data. Whereas integrating IT and OT systems provides promising solutions in a plethora of ubiquitous industrial ecosystems, it introduces novel attack vectors. Loopholes in the system infrastructure lure adversaries to launch different attacks; for instance, denial of participation, false data injection, data forging, and replay attacks, to name a few~\cite{suhail2020role}. 
Such attacks can have catastrophic consequences for the high-assurance applications that are involved in crucial decision-making processes based on sensor data~\cite{neshenko2019demystifying}. 
Thus, to provide protection against data forgery, potential countermeasures for data security highlight the need for rigorous scrutiny to thwart vulnerabilities, enforce verifiable capabilities, and ensure integrity.

% How to secure data...
Considering the involvement of multiple entangled participating entities in intricate industrial processes, the overarching question is:
~\emph{How to ensure the trustworthiness of data collated from disparate data repositories?}
In this regard, blockchain technology gained ground to address challenging issues in the industry with regard to product lifecycle data management and data security~\cite{suhail2020trustworthy}. Utilizing blockchain allows industries to manage data on a shared distributed ledger to solve important glitches in traceability, and record events in a secure, immutable, and irrevocable way~\cite{deloitte}. Relying on the trustworthy data from the blockchain and inputting the retrieved data to a data-driven system to draw intelligent information can certainly open ways for the industries that consider several factors for maximizing their revenues, such as risk prognosis, cost management, and quality assurance~\cite{suhail2019orchestrating}.

For predictive maintenance, one of the promising solutions is to create a 
digital fingerprint (or virtual replica) of the underlying product, process, or service. Such a replica can be used to analyze, predict, and optimize all operations before carrying out its real-world implementation.
Following a closed-loop, the simulation data are fed back to the physical 
% system
object to calibrate the operations and to enhance the system performance. Such a bi-directional mapping between~\emph{physical space} and~\emph{virtual space} is called~\emph{Digital Twin} (DT)~\cite{tao2018digital}. As physical assets begin to operate, DTs collect and integrate data from multiple sources, such as sensor data from factory floor, historical production data acquired from product lifecycle data, and domain knowledge to generate comprehensive data in the form of models, simulations, replications, or behavioral analytics. During industrial processes, DTs run synchronously with their physical counterparts where the primary objective is to track and trace~\emph{data inconsistencies} between the physical and virtual objects. Data inconsistencies between the physical object and virtual object call for the adoption of better calibration and testing strategies that evolve DT models and physical counterparts to support problem-solving in manufacturing, more accurate estimation, prediction, and optimization of the industrial processes~\cite{tao2017digital}.

% What can be achieved by combining DTs and BC...
Combining DTs and blockchain can reshape the industry such that blockchain ensures secure data management and DTs use trustworthy data as input to extract actionable insights for predictive maintenance. For instance, in a manufacturing plant for asset management, on the one hand, upon utilizing a trust architecture (such as~\cite{dedeoglu2019trust}) for sensor data cross-validation, trustworthy key threshold data from the sensors and actuators (affixed with equipment) can be stored and retrieved from the blockchain and can be used to detect trends (such as excessive vibration or heat) to avoid failures or operator injury. On the other hand, through DTs, maintenance and operations personnel can acquire more thorough and accurate insight into asset performance based on the analytics and cognitive data generated on the factory floor. Also, other regulators and suppliers of plant equipment can monitor the data logged at blockchain for preventive maintenance, conduct timely inspections to ensure equipment reliability in addition to record their work or analysis on the blockchain ledger~\cite{miller2018blockchain}.

In this survey, we try to answer the following research questions (RQs.):
\begin{enumerate}
\item [RQ1.] Do we really need blockchain in DTs?
% \item [RQ1.] What is the significance of blockchain in DTs?
% RQ1. How can DT data be shared among multiple untrusted lifecycle parties while ensuring confidentiality, integrity, and availability?
\item [RQ2.] What are the current research trends of DT-, blockchain-, and security-related design solutions?
% \item [RQ2.] What are the existing DT-, blockchain-, and security-related design solutions? 
% \item [RQ3.] What is the role of Artificial Intelligence (AI) in blockchain-based DTs?
\item [RQ3.] How AI will set a course for blockchain-based DTs?
\item [RQ4.] What are the open research and deployment challenges with blockchain-based DTs?
\end{enumerate}

\subsection{Existing Works}
To the best of our knowledge, no survey has been conducted that thoroughly investigates design and implementation issues in blockchain-based DTs. However, some of the recent work emphasized the integration of blockchains for DTs. Starting with the most relevant paper, in~\cite{yaqoob2020blockchain}, the authors taxonomize DTs while providing insights into DTs levels, design phases, enabling technologies, and their potential in core applications. The authors explore the benefits of using blockchains for DTs and discuss DTs synergies and case studies for the automobile industry including Genie, Volvo, and Volkswagen. This study outline technical and social barriers and identify open issues to the adoption of blockchain in DTs, for instance, scalability, data privacy, limitations due to resource-constrained devices, and interoperability issues due to heterogeneous consensus models, transaction schemes, and smart contracts. Also, it outlines security features such as confidentiality and integrity to protect DTs design and operational information, however does not consider other security-related design details such as access control and quantum immunity. A more recent theoretical work~\cite{RAJ2021267} covers applications and benefits of DTs in different industry verticals including manufacturing, automotive, logistics, utilities, aircraft, and healthcare. The focal point of this work is to create blockchain-based DTs to facilitate digital identity and data tracking through blockchain traceability. Besides, it also highlights other features of blockchain including decentralization, safe and secure data transfer, data immutability, authentication of users and data sources in connection with DTs industrial use cases. Despite providing a succinct overview of the blockchain-based DTs, these works do not cover the existing literature on blockchain-based DTs approaches. 

In~\cite{gotz2020exploring}, the authors explore the interoperability and integrability of blockchains and DTs
% blockchain-based DTs
for asset life cycle management. They proposed a model of the blockchain-based DT framework based on surveys of existing literature and questionnaire to industry professionals and mapping of technological use cases for DTs, IoT, blockchains, and smart contracts. This work also outlines integration constraints of blockchain-based DTs with possible solutions. For instance, focus on separable blockchains for easy troubleshooting and combating errors arising during scaling networks and collaboration of the different system areas. Nevertheless, the paper does not survey works on blockchain-based DTs in particular. Similarly, a blockchain-oriented work~\cite{leng2020blockchain} focuses more on the role of blockchain in manufacturing with fewer insights into DTs, whereas a DT-oriented study~\cite{rasheed2020digital} merely touches blockchain technology as a key to solve trust in information issues.

By narrowing down our survey to blockchain-based DTs featuring industrial use cases and focusing on more high-level issues, we present a holistic approach towards blockchains for DTs while presenting design components, fusion with enabling technologies, benefits, and challenges. 
% Research gaps in existing work: 
To this end, we also cover the existing works based on the proposed design and development strategies through a survey of blockchain-based DTs.  We primarily focus on the existing literature that covers the design and implementation aspects of blockchain-based DTs. We choose manifold criteria that cover different design aspects related to DTs, blockchain, and security. In addition to comparison of current solutions, we also emphasize challenges that are yet to be addressed.

\subsection{Scope of This Survey}
In this survey, we present a thoroughly comprehensive, and systematic review of state-of-the-art results for blockchain-based DTs. We also focus on blockchain-based DTs design and implementation issues in detail and provide possible solutions to address such issues. Additionally, we investigate technical, logistical, and social issues that arise due to the integration of blockchain with DTs and must be addressed by stakeholders. We summarize the main contributions of this paper below:

\begin{itemize}
    \item We present an in-depth comprehensive and systematic review of the state-of-the-art blockchain-based DTs approaches covering design and implementation issues in terms of blockchain, DTs, and security. 
    \item Based on the limitations of the current works, we propose a framework for trustworthy blockchain-based DTs. 
    \item In addition to ongoing standardization efforts, we provide an in-depth review of various research challenges such as technical, logistical, and social challenges. In contrast to existing surveys, we also map such requirements from both blockchain and DTs perspectives, provide pragmatic approaches for the adoption of blockchain-based DTs, and highlight the open-ended challenges that need to be addressed by the industry and research community. 
\end{itemize}  

The rest of the paper is organized as follows: Section~\ref{background} covers a quick high-level overview of DTs, blockchain, and the major benefits of using blockchain for DTs. Considering blockchain, DTs, and security as the key factors, we present the design and implementation details of existing research in Section~\ref{research-trends}. Section~\ref{DT_AI} discusses the role of enabling technologies in blockchain-based DTs. Section~\ref{challenges} describes the technical, logistical, and social challenges and requirements of blockchain-based DTs. Finally, Section~\ref{conclusion} concludes the paper with takeaways from this study.

\section{Background} \label{background}
This section outlines the necessary background information for this survey. We first provide the main design components and applications of DTs in the industry followed by a discussion on blockchain and its role in ensuring secure data management in the industry.
\subsection{Digital Twins}
\begin{figure}[!ht]
 \centerline{\includegraphics[width=3.0in]{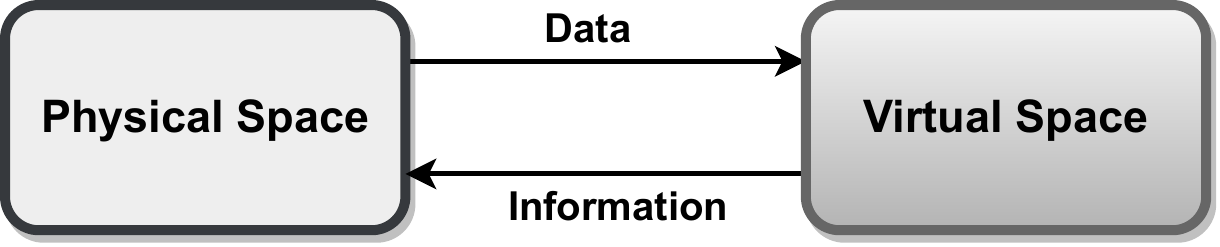}}
\caption{Adaption of Grieves' 3D DT model~\cite{grieves2014}.}
\label{fig:3Dmodel}
\end{figure}

% Evolution of DTs: DTs 3D and 5D model 
The concept of DT was first introduced by Micheal Grieves in 2003 during his course on Product Lifecycle Management (PLM). In 2012, the National Aeronautics and Space Administration (NASA) formalized the definition of DTs as an integrated multi-physics, multi-scale, probabilistic simulation mirroring the physical system based on models, sensors, history data, etc.~\cite{glaessgen2012digital} in the field of the aerospace industry. In 2014, Grieves' white paper~\cite{grieves2014} proposed a preliminary form of a three-dimensional (3D) DT model that includes a physical space containing a physical entity, a virtual space containing a virtual entity, and a connection to allow synchronization of data between both spaces (as shown in Fig.~\ref{fig:3Dmodel}). In continuation of the role of DTs in the aviation industry, DTs lead to the notion of Airframe Digital Twin (ADT) for predicting the life of aircraft structure and maintenance needs~\cite{tuegel2011reengineering}. In parallel, continuous advances in DT enabling technologies, for instance, IoT, big data processing techniques, cloud computing, and AI further motivate the adoption of DTs beyond the aerospace industry and thus are promoted in numerous applications including manufacturing, supply chain, smart cities, industrial IoT, servitization, healthcare, etc. According to Gartner's survey report, $13\%$ of enterprises are already using DTs, whereas $62\%$ are in the process of deploying DTs~\cite{Gartner}. 

\begin{figure}[!ht]
 \centerline{\includegraphics[width=3.0in]{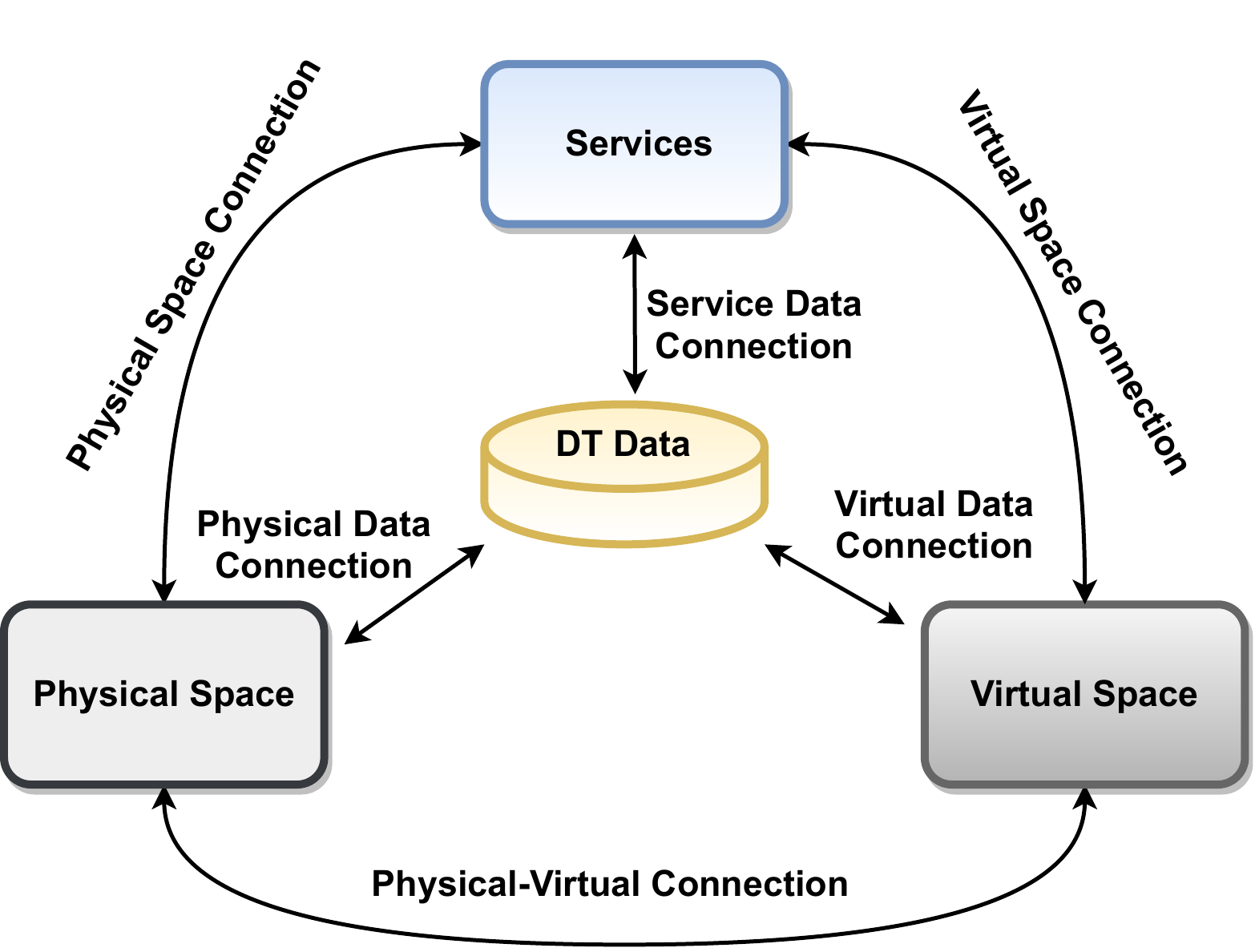}}
\caption{Adaption of 5D DT model~\cite{tao2018digitalmodel}.}
\label{fig:5Dmodel}
\end{figure}

Based on the Grieves' 3D DT model, Tao et al. proposed a five-dimensional (5D) model that includes: physical space, virtual space, connection, DT data, and service (as shown in Fig.~\ref{fig:5Dmodel})~\cite{tao2018digitalmodel}. The physical part supports the virtual part through assets data while the virtual part controls the physical part through simulation and decision-making. Service supports the management of both spaces during operation, evolution, and optimization. DT data repository acts as a primary source for knowledge creation and connection bridges all parts of the DT model.

% Role of DTs in industry 
The data-driven DTs focus on the areas of product design, production, manufacturing, operational control, Prognostic Health Management (PHM), etc. in the industrial domain~\cite{tao2018digitalsurvey}. For instance, DTs enable more accurate predictions, rational decisions, and informed plans by reinforcing the collaboration between design and manufacturing phases. PHM plays a significant role in the lifecycle monitoring of equipment conditions where a DT of the equipment supported by data is used to depict the erratic behavior of the physical object, perform fault diagnosis, and provide design rules for maintenance~\cite{gao2015cloud}. Similarly, dispatching optimization and operational control are among other promising application areas of DTs~\cite{tao2018digitalsurvey}. Industry leaders including International Business Machines Corporation (IBM), Siemens, General Electric (GE), Microsoft Azure, Airbus, Systems, Applications and Products in Data Processing (SAPSE) have deployed DTs in various industrial application areas.

\subsection{Blockchain}
The emerging technology of blockchain has attracted immense attention from academics and practitioners in diverse disciplines including finance, law, and computer science due to its promising secure distributed framework that facilitates sharing and synchronization of information across all participating entities~\cite{puthal2018blockchain}. In the blockchain, transactions among nodes serve as the basic communication primitives that are recorded as append-only time-stamped logs in a shared digital ledger. To establish a secure trusted network among untrusted nodes, a consensus mechanism is employed that involves solving a resource-demanding hard-to-solve and easy-to-verify cryptographic puzzle. Following the consensus algorithm, transactions are collated to form a block which is then appended to the ledger. To enforce immutability in blockchain structure, each block contains the hash of the previous block, thereby can readily detect tampering to a block and hence transactions. 

The rapid adoption of blockchain has largely been spurred by the success of Bitcoin (the first prototype of cryptocurrency)~\cite{nakamoto2019bitcoin} and successfully replaced economic transaction systems in many organizations. Other alternatives (Ethereum and Ripple) continue to mature, thus proving to be a highly disruptive force in finance and e-commerce sectors~\cite{henry2018blockchain}. Later on, numerous non-monetary application scenarios are identified as potential candidates of blockchain-based solutions, for instance, intellectual property, proof of location, voting, healthcare, to name a few. Recently, blockchain-based solutions in combination with IoT are gaining ground to revamp heterogeneous business models (such as Business to Business (B2B) and Business to Consumer (B2C)) in a variety of industrial sectors, including supply chain and manufacturing plant asset management~\cite{miller2018blockchain}.

\begin{figure}[!ht]
 \centerline{\includegraphics[width=3.5in]{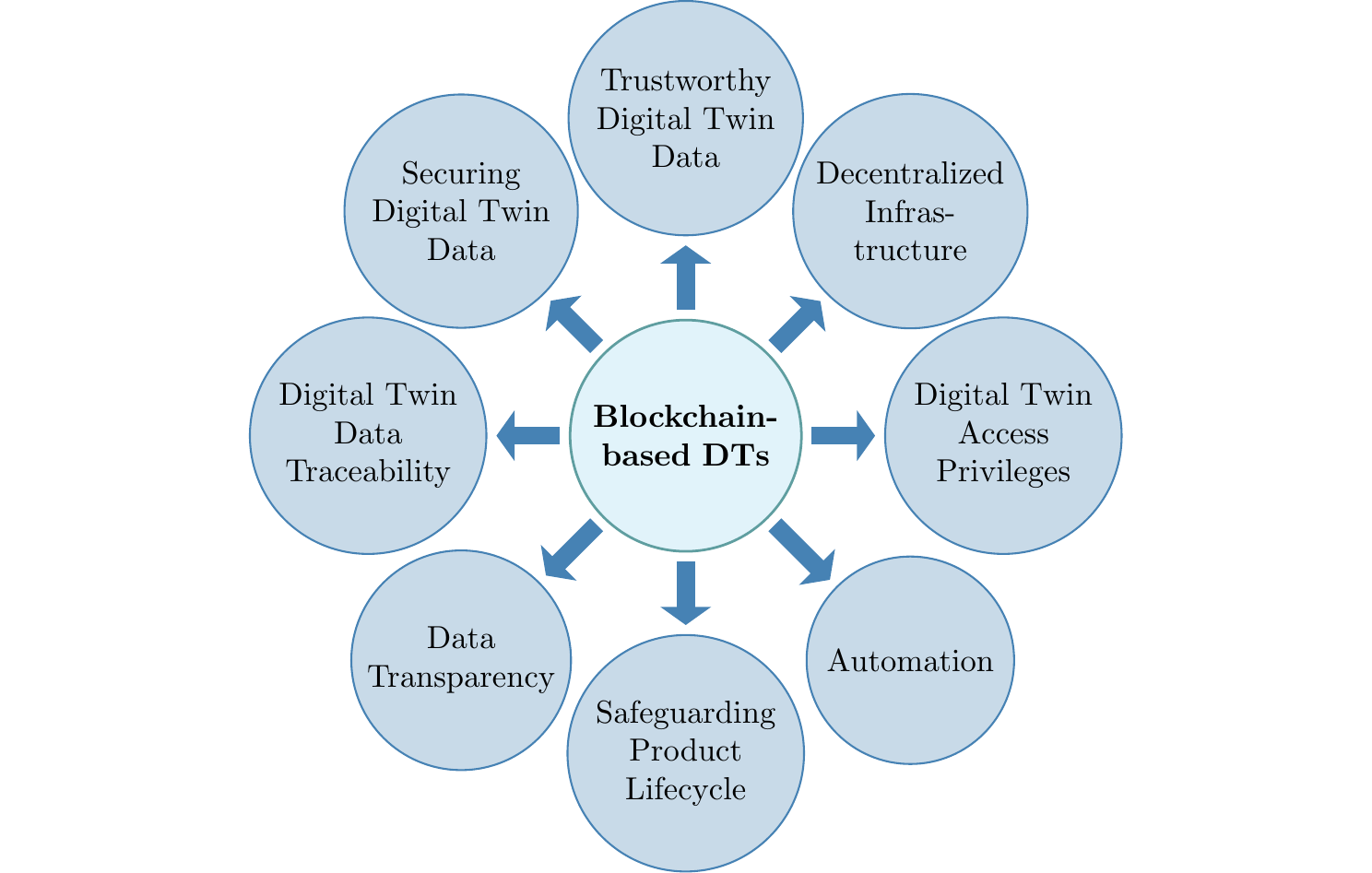}}
\caption{Benefits of using blockchains for digital twins.}
\label{fig:benefits}
\end{figure}

\subsection{Major Benefits of Using Blockchain-based Digital Twins in Industry} \label{benefits}

% What is the PROBLEM?
Although DTs enable the realization of futuristic services and applications of industry in a transparent and efficient way, these benefits of DT are based on an assumption about data trust, integrity, and security. Nevertheless, in real-life scenarios, data breaches could occur due to a number of reasons, both malicious and otherwise. Therefore, to fully harness the features of DT, the data used by the DTs needs to be trustworthy and secure. 
% Solution
In this regard, leveraging blockchain technology allows industries to manage data on a distributed ledger while ensuring trusted DTs data coordination.
% Do we really need BC for DTs
Nevertheless, we need to critically analyze the question,~\emph{do we really need blockchain for DTs?} based on the assessment elements including the number of readers/writers of the data, presence of a trusted third party, identification of readers/writers, and the nature of verifiability of the exchanged data~\cite{wust2018you}. According to the model of Wust et al.~\cite{wust2018you}, a permissioned blockchain can be the right candidate for DT to cope with these issues because of, but not limited to, the cryptographic primitives of blockchain guarantee data immutability and the consensus mechanism of the blockchain (where different stakeholders of the data come to a common consensus while managing the data) provides transparency and accountability. In the following, we discuss the key benefits (summarized in Fig.~\ref{fig:benefits}) of using blockchain-based DTs.

% /////
\subsubsection{Trustworthy Sources}

% What is the PROBLEM?
The physical-digital mapping in the industrial processes is achieved through IoT-based technologies that might be vulnerable to cyber threats~\cite{suhail2021securing}. Since, the accumulated data accounts for rational decision making, therefore, it is important to manifest the trustworthiness of data-generating sources.
% Role of blockchain:
Blockchain can play a significant role by collecting data from registered devices to proactively circumvent device tampering or utilizing a blockchain-based layered trust architecture~\cite{dedeoglu2019trust} for cross-validation of sensor data. A more detailed discussion is provided in Subsection~\ref{subsec:trustworthysources}.

% /////
\subsubsection{Distributed Infrastructure}

% Requirement of DTs:
The lifecycle of DT span over different phases where multiple stakeholders (may be geographically dispersed) carry out various operations on it. The centralized data management architectures are not well suited for such environments. Additionally, the disparate data silos act as a barrier in orchestrating the DT lifecycle particularly in digital thread that requires linking data throughout various phases of the lifecycle.
% Solution from Blockchain:
To enable the storage of information
in a distributed manner with no single point of control can be established through blockchain.  
Thus, not only the DTs' data but data from other sources (such as data generated within the production facility or during the supply chain trade events), provenance data, etc. flowing across multiple tiers or entities can be stored and accessed through the blockchain ledger~\cite{suhail2020trustworthy}.

% /////
% \subsubsection{Secure Digital Twins Data}
\subsubsection{Securing Digital Twins Data} \label{subsec:securingDTdata}

% What is the PROBLEM?
The increased digitization and connectivity opens up new attack vectors that may put industry assets or process at risk whereas DTs are required to be built on trusted data. The question how to provide integrity and trustworthiness of data collected from the physical world that are then shared with the digital world needs to be answered. 
% Role of Blockchain
The tamper-proof and immutable nature of blockchain can be leveraged to procure data across multiple participating entities in a secure manner. For instance, the cryptographic strength blockchain keeps the irrevocable history of DTs creation or access transactions. Furthermore, the use of pseudonyms, such as public keys for representing node identities, for blockchain transactions can support the coexistence of transparency and privacy. Nodes can maintain multiple pseudonyms to remain anonymous, to enable confidential data flows, and to cope with industrial espionage for competitive advantage. Thus, DT data can be shared among multiple untrusted (or partially trusted) lifecycle parties while ensuring confidentiality, integrity, and availability.

% /////
\subsubsection{Data Traceability in Digital Twins}

Additionally, provenance-enabled blockchain-based DTs can reason about the current state of a data object such as~\textit{why},~\textit{where}, and~\textit{how} throughout the process chain~\cite{suhail2020trustworthy}. The complete lineage of DTs ranging from their origin to the present can be accessed to identify the cause of deviations between the physical and digital space in addition to data assessment (such as data debugging, reconciliation, performance tuning, auditing, etc.).

% /////
\subsubsection{Access Privileges of DTs}
In the case of the enterprise industry, both reading and modification to the data are subject to access rights. This argument can be true for DT as well where the data can be viewed by the selected set of entities and the modifications must be done by only authorized entities. Blockchain-based access control manages policies and rules defined for DTs. Moreover, blockchain inherently retains the history of modifications and thus can circumvent the problem of illegal data modification that may lead to other data security-related problems.

% /////
% \subsection{Safeguarding Digital Twin Instance}
\subsubsection{Safeguarding Product Lifecycle Data}\label{subsec:DTinstances}

To facilitate an abstract view of overall phenomena, DTs must support the correlation of different DTs associated with different physical sub-components~\cite{suhail2021securing}. For example, Digital Twin Instance (DTI) is a singular entity that mirrors the physical asset and/or activity, whereas Digital Twin Aggregate (DTA) is the combination of DTIs (preferably similar DTIs) to mimic the bigger picture of the physical world~\cite{RAJ2021267}. The DTA represents an attractive target for attacks and may lead to severe consequences as the whole product line can be exposed to data breach (for example, manipulated updates to put the DT into a malicious state). Blockchain-empowered DTs provide adequate security measures to protect DTIs within the DTA by allowing only authorized  entities to make changes in addition to data integrity and trust as discussed in~\ref{subsec:securingDTdata}.

% /////
\subsubsection{Automation Through Smart Contract}
The industrial processes demand automation to trigger timely and seamless actions. For instance, logging data in the course of change in physical state and its counterpart. To enforce automation, smart contracts from blockchain can be deployed to automate scenarios depending on the underlying application requirements. Subsection~\ref{subsec:smartcontract} presents more examples on this.

\subsubsection{Combating the Problem of Counterfeits}
Creating counterfeit products that are near-replicas of the original products not only affect the investment of the industry stakeholders but also badly affects the brand value and reputation of the products. To overcome this challenge, the combination of blockchain and DT can ensure that counterfeit products are identified by providing product authenticity. More precisely, the transparency will come from the blockchain (for instance, creating digital certificates on the blockchain) whereas the digitization of the products and assets will be guaranteed by the DT. Therefore, the integration of DT and blockchain will play a pivotal role in ensuring the Return on Investment (RoI) for the industry by removing the problem of counterfeits. Additionally, blockchain-based solutions for DTs can provide verifiable digital certificates to solve important glitches regarding the legitimacy of DTs by identifying counterfeiters attempting to imitate or distribute fake replicas of DTs~\cite{RAJ2021267}.

% /////////////////////

\section{Blockchain-based Digital Twins: Research Trends}\label{research-trends}
\begin{figure}[!ht]
\centerline{\includegraphics[width=3.0in]{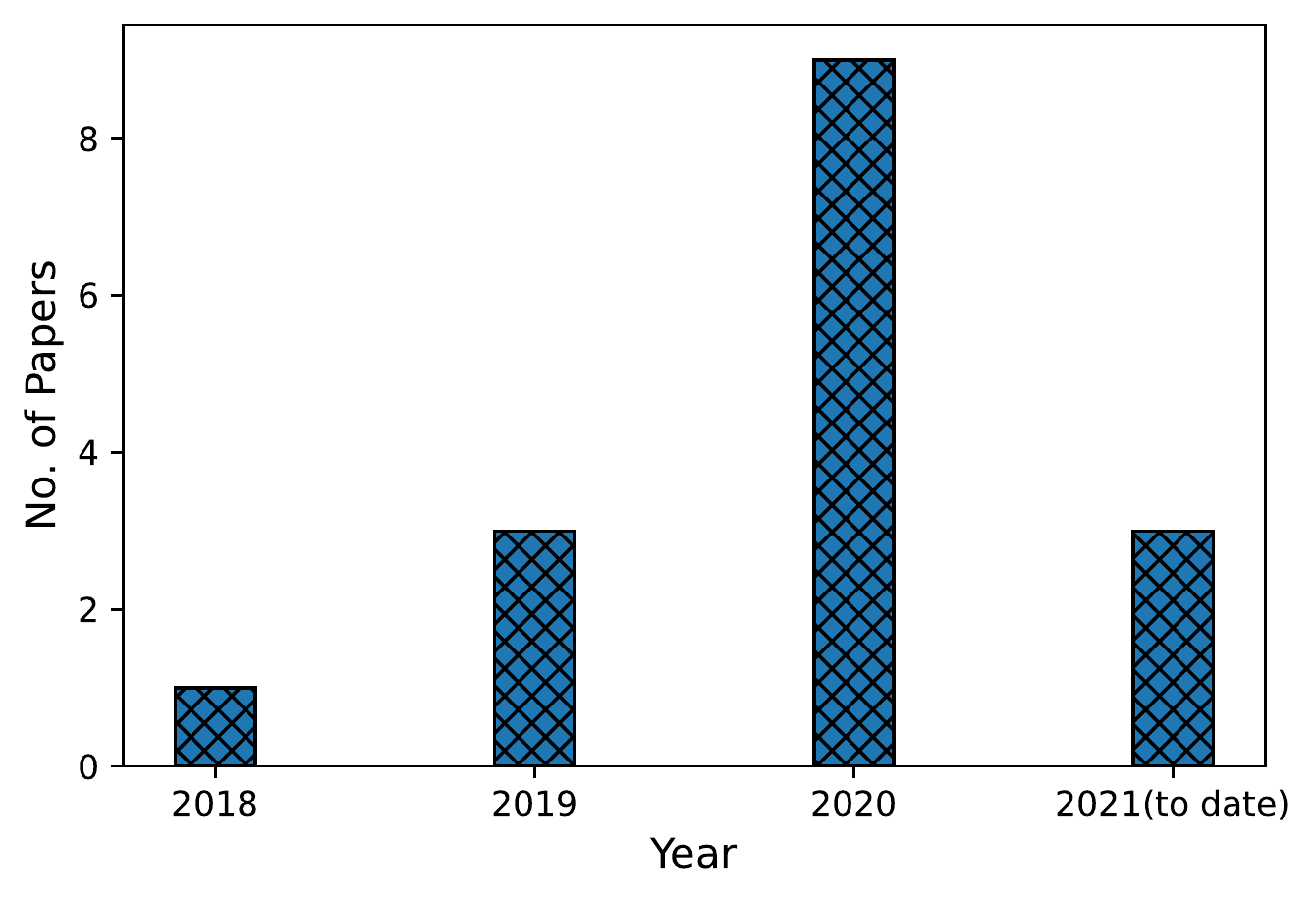}}
\caption{Blockchain-based DTs in industry: Research trends.}
\label{fig:trends}
\end{figure}

\begin{figure}
    \centering
    \centerline{\includegraphics[width=7.3in]{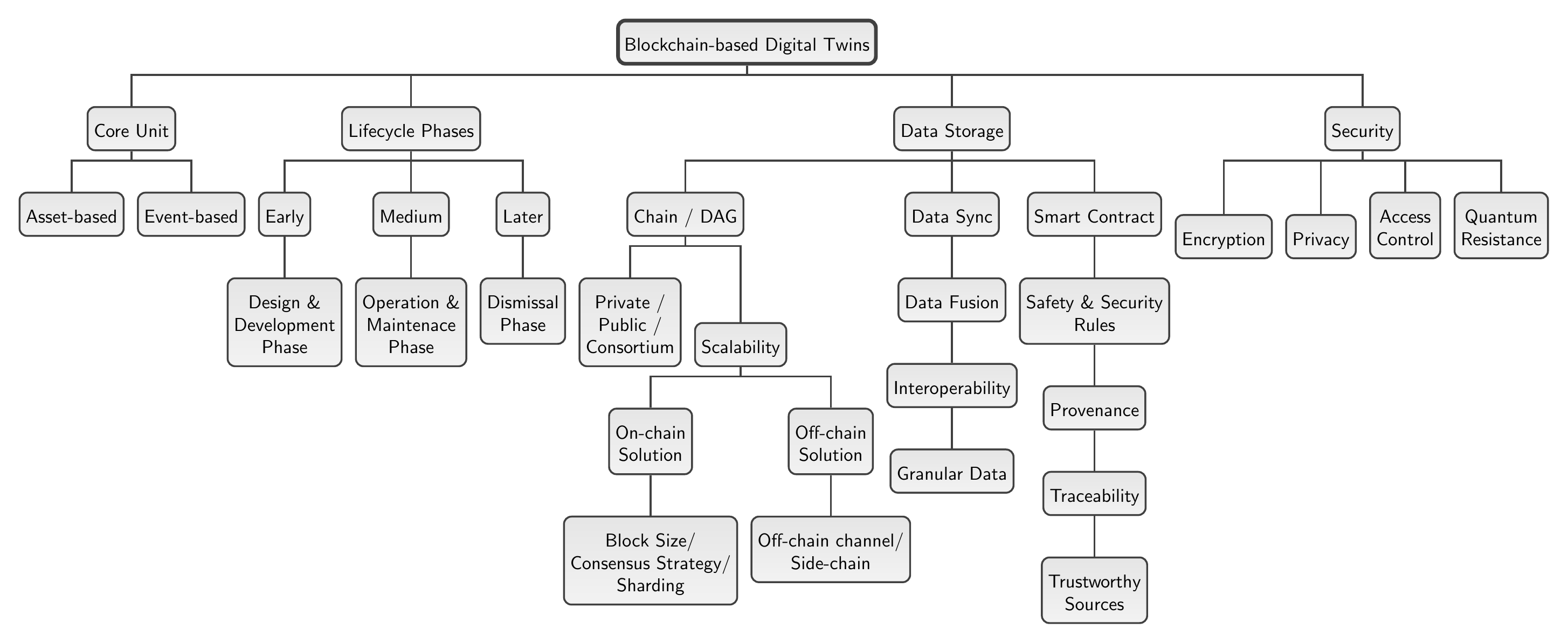}}
\caption{Taxonomy: Blockchain-based digital twins}
    \label{fig:taxonomy}
\end{figure}

In this section, we review the existing works that focus on the design and implementation aspects of blockchain-based DTs. To discuss the existing works, we choose manifold criteria that cover different design aspects related to DTs, blockchain, and security. In addition to discussion on current solutions, we also emphasize challenges that are yet to be addressed. Furthermore, we provide a detailed discussion on a trustworthy blockchain-based DT that tries to address some of the outstanding challenges in the existing works. Fig.~\ref{fig:trends} shows the research trends of exploring the integration of blockchain with DTs by the research community. It can be clearly seen from Fig.~\ref{fig:trends} that there is an increased interest in the integration of blockchain with DT. Fig.~\ref{fig:taxonomy} depicts the taxonomy which is devised based on the design solutions for blockchain-based DTs in industry.
Table~\ref{tab:review_technicalwork} comprehensively summarizes relevant recent works on blockchain-based DTs. This table shows three main interesting design considerations for DT, blockchain, and security from the blockchain-based DT literature. Firstly, it covers DT-related design details including DT definition, component composability and integration, lifecycle phases, and implementation aspects. Secondly, it outlines blockchain-related design details including the type of blockchain in the proposed solution, implementation aspects, off-line data storage mechanism, traceability, and trustworthy data sources. Thirdly, it highlights the security-related design details including encryption, access control, and quantum resilience. In the following, we discuss each of them in detail. 

% /////////////////////
\begin{landscape}
\thispagestyle{empty}
 \renewcommand*{\arraystretch}{3.0}
\begin{table}[!ht]
\caption{Comparison of blockchain-based digital twin approaches.}\label{tab:review_technicalwork}
\small
\tabcolsep=0.075cm
\centering
\begin{tabular}{@{}*{21}{c}@{}}

\toprule
\hline
\multirow{3}{*}{\thead{\textbf{Ref.}}}  & 
  \multicolumn{9}{c}{\thead{\textbf{Digital Twin-related} \\ \textbf{ Design Details}}} &
  \multicolumn{8}{c}{\thead{\textbf{Blockchain-related} \\ \textbf{ Design Details}}} &
  \multicolumn{3}{c}{\thead{\textbf{Security-related} \\ \textbf{ Design Details}}} \\
  \cline{2-21} 
% \midrule
 & 
  \multirow{2}{*}{\thead{\it Defin-\\\it ition}} &
  \multirow{2}{*}{\thead{\it Compo-\\\it nent}} &
  \multirow{2}{*}{\thead{\it Lifecycle\\ \it Phases }} &
  \multicolumn{6}{c}{\it Implementation} &
 \multicolumn{5}{c}{\it {Implementation}} &
  \multirow{2}{*}{\thead{\it Data\\ \it Storage}} &
   \multirow{2}{*}{\thead{\it Trace-\\\it ability}} &
   \multirow{2}{*}{\thead{\it Trust-\\\it worthy\\ \it Sources}} &
  \multirow{2}{*}{\thead{\it Encry-\\ \it ption}} &
   \multirow{2}{*}{\thead{\it Access\\\it Control}} &
  \multirow{2}{*}{\thead{\it Quantum\\\it Resilient}} \\ \cline{5-15} 
    & 
    &
   &
   &
     \thead{\it Data\\\it Fusion} &
  \thead{\it Inter-\\\it operable} &
  \thead{\it{S\&S}\\\it Rules} &
  \thead{\it Data\\\it Sync} &
  \thead{\it Granular\\\it Data}    &
  \thead{\it {Inter}\\\it {face}}    &
   {\it Type} &
  \thead{\it {Scala-}\\\it {bility}} &
  \thead{\it {Smart}\\\it {Contract}} &
  \thead{\it {Consensus}\\\it {Protocol}} &
  \thead{\it Proto-\\\it type} &
    &
   &
   &
   &
 \\
%   \hline
\midrule
   %   //////////////////////
    ~\cite{angrish2018case} & \thead{machine\\ events} & \xmark & M & \xmark & \xmark & \xmark & \xmark & \xmark &\xmark & ETH & off-chain  & \cmark  & \thead{PoW\\PoA} & \faAdjust &  \faAdjust  &\xmark & \xmark & \xmark  & \faAdjust & \xmark 
    \\ \hline
% % %   //////////////////////
% %   %   A distributed ledger approach to DT secure data sharing 2019
  ~\cite{dietz2019distributed} & any asset & \xmark & E\&M & \xmark & \faAdjust & \faAdjust & \faAdjust  & \xmark & \cmark & ? &  off-chain & \cmark  & \xmark & \xmark & \faAdjust &  \faAdjust & \xmark  &  \faAdjust & \faAdjust & \xmark  
  \\ \hline
% % %   //////////////////////
% % % Building a digital twin for additive manufacturing through the exploitation of blockchain: A case analysis of the aircraft industry 2019
~\cite{mandolla2019building} & product & \xmark & E\&M & \xmark & \xmark & \xmark & \xmark &  \xmark  & \xmark & ? & \xmark & \xmark& \xmark  & \faAdjust & \xmark & \xmark & \xmark & \xmark & \xmark & \xmark
  \\ \hline
% % %   //////////////////////  
% %   %MBCoT 2020
  ~\cite{zhang2020manufacturing} 
  & \thead{manufac\\turing\\ events} & \xmark &  E\&M & \xmark  & \faAdjust & \xmark & \xmark &  \xmark & \xmark & HLF & off-chain & \cmark & XFT & \faAdjust & \faAdjust & \faAdjust & \xmark & \xmark & \faAdjust & \xmark
  \\ \hline
% % % %   /////////////////////
% % %  Blockchain-based data management for digital twin of product 2020
~\cite{huang2020blockchain}
& product & \xmark & E\&M & \xmark & \xmark & \xmark & \faAdjust  &  \xmark & \xmark & ? & \xmark & \cmark & PoW & \faAdjust & \xmark  & \xmark  & \xmark  & \faAdjust & \xmark & \xmark
 \\ \hline
% % %   /////////////////////
% % % A Blockchain-Based Approach for the Creation of Digital Twins 2020
~\cite{hasan2020blockchain} & any asset & \xmark & E\&M & \xmark  & \xmark & \xmark & \xmark &  \xmark & \xmark & ETH & off-chain & \cmark & \xmark & \faAdjust & \faAdjust  & \faAdjust  & \xmark & \xmark & \xmark & \xmark
 \\ \hline
% % %   /////////////////////
% % % ManuChain: Combining permissioned blockchain with a holistic optimization model as bi-level intelligence for smart manufacturing 2020
~\cite{leng2019manuchain} & \thead{manufac\\turing\\ events} & \xmark & E & \xmark & \xmark & \xmark & \faAdjust &  \xmark &  \xmark & HLF & off-chain & \cmark & XFT & \faAdjust & \faAdjust  & \xmark & \xmark & \xmark  & \xmark  &  \xmark
\\ \hline
% % %   /////////////////////
% % % Towards Smart Manufacturing Using Spiral Digital Twin Framework and Twinchain}, 
~\cite{khan2020towards} & product & \xmark & E & \xmark  & \xmark & \xmark & \xmark &  \xmark & \xmark & variant & \xmark & \xmark & PoW & \xmark & \xmark & \xmark & \xmark & \xmark & \xmark & \cmark
\\ \hline
% % % %   /////////////////////
% % % EtherTwin: Blockchain-based Secure Digital Twin Information Management 2021
~\cite{putz58ethertwin} & any asset  & \xmark & L & \xmark & \xmark & \xmark & \faAdjust &  \xmark & \cmark & ETH & off-chain & \cmark & \xmark & \cmark & \cmark & \xmark & \xmark & \cmark & \cmark & \xmark
\\ \hline
% %   //////////////////////////////////////////////////////
\bottomrule
\end{tabular}
 Legend: \textbf{?}~Unknown; \xmark~Not covered; \faAdjust~Partially covered; \cmark~Covered; \textbf{E}~Early; \textbf{M}~Medium, \textbf{L}~Later.
 
\end{table}
\end{landscape}
% /////////////////////

% /////////////////////
\subsection{Design Solutions for Digital Twins}
% For each col. 
% Current Solutions
% Challenges
% How to connect it with above para....
% ///// Def. + Component
% How (Existing Solutions) mention DT def and components (col. 1 and 2)
The majority of existing works consider DTs as a representation of any asset, product, machine, or manufacturing events. However, the existing works do not cover the integration and interoperability of multiple components or sub-components to provide an aggregated view of the complex physical system. 
% Link to DTI and DTAs
For instance, combining DTIs of the physical asset and/or activity to form DTA to generate the bigger picture of the physical world. 
% (Challenges)
As a result, the lack of correlation among different elementary DTs leads to the failure of abstract views of larger phenomena. In the following, we discuss the design solutions for DTs from different perspectives such as design lifecycle, fusion and interoperability, security and safety, and synchronization and consistency. These are the cornerstones of DT design and we discuss both the existing works in these directions and the outstanding challenges.

\begin{figure}[!ht]
 \centerline{\includegraphics[width=3.5in]{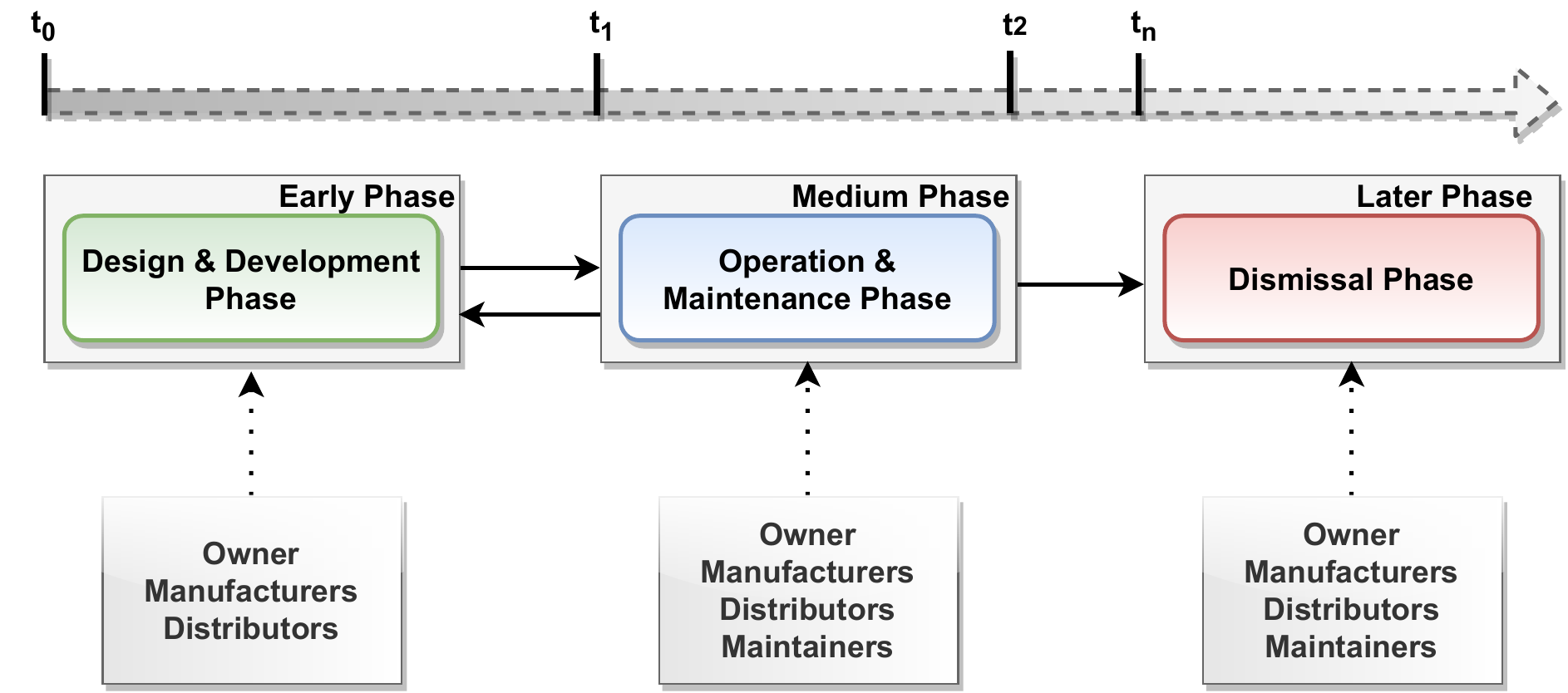}}
\caption{Lifecycle of digital twins: From design to dismissal.}
\label{fig:lifecycle}
\end{figure}

\subsubsection{Digital Twin Lifecycle}
Fig.~\ref{fig:lifecycle} represent a DTI where the data linkage throughout various phases of the lifecycle can be performed through digital thread.
The lifecycle phases of a DTI start from design to dismissal, in addition to the potentially involved lifecycle parities (i.e., owner, manufacturers, distributors, and maintainers)~\cite{barricelli2019survey}. Firstly, the asset prototype is designed and is evolved based on historical data, static data, and simulation data during the {\it design phase}. Then the physical asset is tailored based on the asset prototype and other technical requirements during the {\it development phase}. Secondly, following a synergic and continuous interaction between the prototype and the asset, the asset further evolves and optimizes its functionalities during {\it operational and maintenance phase} while the prototype grows and self-adapts to the asset during its existence. Finally, due to obsolescence or other reasons, the asset along with its DTs is demolished during the {\it dismissal phase}. 
% How (Existing Solutions) mention lifecycle phases 
The existing works have usually focused on tackling early (~\cite{leng2019manuchain,khan2020towards}), medium (~\cite{angrish2018case}), or both of these phases (~\cite{dietz2019distributed,mandolla2019building,zhang2020manufacturing,huang2020blockchain,hasan2020blockchain}) of the DTs lifecycle. 
% What is missing in existing works (Challenges)
While developing a prototypical blockchain-based DT solution called {\it EtherTwin}, Putz et al.~\cite{putz58ethertwin} mentioned the dismissal phase during the DT lifecycle; however, they do not discuss the details such as how the object is disassembled and the stored historical data of the DT product are backed-up and made available to other objects or domain experts to optimize the production of future devices. 
% //////

\subsubsection{Data Fusion and Interoperability}
Regarding the implementation details of DTs, fusion and interoperability of data from multimodal and heterogeneous sensors is one of the significant prerequisites. This involves cleaning invalid, duplicate, or missing data, converting heterogeneous data formats into a unified data structure, and then fusion of such data to generate a consistent interpretation of a certain object.
% How (Existing Solutions) mention Data Fusion and Inter-operable
% What is missing in existing works (Challenges)
As depicted from Table~\ref{tab:review_technicalwork}, none of the existing works consider the process of data fusion from technical perspective. Similarly, interoperability is partially addressed by~\cite{zhang2020manufacturing, dietz2019distributed}. For example, a hardware infrastructure proposed by Zhang et al.~\cite{zhang2020manufacturing}, introduced adopters for parsing the data collected from systems with different communication protocols. Additionally, it includes edge gateway nodes to pre-process the perceived data through filtration and aggregation. Similarly, Dietz et al. \cite{dietz2019distributed} discussed the integration of data in multiple formats to support data variety. They categorized the data as semi-structured data to model physical asset (static properties of the device) by using AutomationML (AML) format, structured data (sensor data), and unstructured data (human-readable documents). 
% however, still lack technical facets. 
However, the above-mentioned theoretical solutions lack implementation details to demonstrate their feasibility in practice.
% Ending...
% Interoperability allows data fusion that provides a more
Interoperability and data fusion provides a more
comprehensive, consistent and accurate representation compared with the single perspective of data. However, the existing works still lack due attention to the relation of data fusion with the DT functionality.

% First write why we need SS and what are the implications if they are missing... 
\subsubsection{Safety and Security (S\&S)} \label{sec:SS}
Some other implementation aspects including Safety and Security (S\&S) threats, malicious or accidental, must be integrated to detect misconfigurations and to mitigate the outcomes of malfunctioning components and malicious activities in a production system. For example, safety rules may define upper- and lower-speed thresholds on a motor~\cite{eckhart2018towards}. Inability to impose such rules leads to devastating results on the system in the form of cyberattacks or Advanced Persistent Threats (APT), for instance, Stuxnet worm~\cite{langner2013kill} and Triton~\cite{miller2019triton}. 
% How (Existing Solutions) mention SS
% What is missing in existing works (Challenges)
Despite being a critical construction parameter of~\enquote{security by design}, S\&S rules are overlooked by existing works. To the best of our knowledge, only one of the existing works by Dietz et al.~\cite{dietz2019distributed}, suggested adopting S\&S works by referring to interesting work by Eckhart and Ekelhart~\cite{eckhart2018towards}. In this work, the authors consider monitoring of S\&S rules and demonstrated the detection of the notorious Man-in-The-Middle attack (MiTM) attack by targeting the manipulation of a motor's speed. Moreover, to further explore this parameter, future directions such as modeling language for S\&S rules, investigating device templates with security rules, behavior learning, and analysis 
% in combination
in conjunction with anomaly detection are also highlighted. Introducing S\&S mechanisms at the DT design phase can help to lower security and incident-response costs thereby making later lifecycle phases less prone to errors and incidents~\cite{dietz2020unleashing}. Whereas, S\&S rules must be stored and retrieved from the ledger to ensure their reliability. This further strengthens the rationale for integrating blockchain with DTs.

\subsubsection{Synchronization and Consistency}
Another key factor of DT-related design is the synchronization and consistency of data between the physical and digital spaces. 
% which are highly dependent 
Both of these factors are highly associated with the volume and velocity of data generation and data storage.
%Now write about how static and dynamic data and then relate it to data volume and velocity.
DT data can be categorized as descriptive (or static) data and behavioral (or dynamic) data. The latter changes infrequently with time along the lifecycle of the real-world counterpart whereas the former evolves frequently with each lifecycle phase resulting in different versions and a dynamic information structure~\cite{rios2015product}. 
% How (Existing Solutions) mention Data Sync
In the existing works, the issue of supporting data volume and velocity is partially considered.
% Examples from existing works:
For example, Huang et al.~\cite{huang2020blockchain} suggested storing sensor data based on preset time intervals. However, they did not consider the fundamental requirement of real-time data synchronization. On the other hand, Dietz et al.~\cite{dietz2019distributed} discussed a more feasible solution to store the hash reference to the data stream on the ledger while storing data off-chain. Although this solution addresses the data storage issues, the details subjected to performance metrics are still missing. Similarly, Putz et al.~\cite{putz58ethertwin} suggested the off-chain data storage pattern to meet the data volume and latency requirements and suggested sharing only non-nominal or aggregated sensor data to reduce the data volume. 

Leng et al.~\cite{leng2019manuchain} suggested a multi-view synchronization-based DT model that uses blockchain as an indexing server to record the transactions that are critical for the life-cycle quality control and product anti-counterfeiting. They also use traditional databases for the fast retrieval of history transactions.
% What is missing in existing works (Challenges)
Considering the existing solutions, the crux of the problem is that the dynamic data produced by the deployed sensors and actuators providing real-time updates about the asset's actual state usually amass in intervals ranging from minutes to milliseconds. However, sensor data streams' velocity and volume exceed the performance characteristics of mainstream blockchain frameworks.
% distributed ledger frameworks. 
Furthermore, deciding which data to store on-chain and which to off-chain, also needs due attention. 

% Solution
To address data synchronization challenges, the blockchain-based DTs solutions must exhibit high throughput and low sharing latency to efficiently sync data between physical and virtual space in addition to improving data collection and using it intelligently to manage big data challenges. 

\subsubsection{Data Granularity}
Data is considered as the fuel for ensuring precise decision-making while at the same time quality of the data matters at par with quantity for decision-making. Therefore, the amount of collected data holds as an important design parameter.
%  discuss and connect both 
% How (Existing Solutions) mention Data Sync
% What is missing in existing works (Challenges)
The granularity level (fine- and coarse-grained) of collected data which in turn also depends on the optimal number and placing of sensors~\cite{tao2019make}, is still missing %receive no attention by
in the existing works. Further gaps that are required to be addressed include what types of data to sample, which sensors to use and where to install them~\cite{kusiak2017smart}.
% /////% /////% /////End of part I

% Reorganized Section 3.2
\subsection{Design Solutions for Blockchain}
In the following, we discuss blockchain-related design solutions (for instance, selection of blockchain, implementation aspects, on-off chain storage, traceability, and trustworthy data sources) proposed by the existing works and associated challenges.
% Is there a blockchain design that simultaneously scales throughput, storage efficiency, and security?

\subsubsection{Implementation}
%Existing Solutions
While most of the existing schemes proposed only a theoretical blockchain-based framework (~\cite{dietz2019distributed, zhang2020manufacturing, khan2020towards}) neglecting the prototypical implementation. However, several works such as~\cite{mandolla2019building, hasan2020blockchain} only partially accomplished prototype implementation. To date, only Putz et al.~\cite{putz58ethertwin} carried out a thorough implementation. EtherTwin~\cite{putz58ethertwin}, an open-source prototype, is based on blockchain design patterns and DApp technologies with performance and cost measurements. Some other works such as~\cite{huang2020blockchain, mandolla2019building} do exist in the literature but do not provide any technical details about the blockchain type and other evaluation metrics. Similarly, Khan et al.~\cite{khan2020towards} proposed a quantum-immune blockchain variant; however, it lacks technical and other supporting implementation details. On the other hand, in addition to other evaluation metrics, providing a user interface is another important aspect of implementation that is suggested by 
%Dietz et al.  (author names: Marietheres Dietz, Benedikt Putz, and Günther Pernul)
\cite{dietz2019distributed}, however,
in~\cite{putz58ethertwin} they
% Putz et al. (author names: Benedikt Putz *, Marietheres Dietz, Philip Empl, Günther Pernul)
design and additionally implement it.

% Challenges
Overall, the existing works are not extensively evaluated to gain additional practical insights in the industry, and thus require focusing more on the prototype to identify potential adoption barriers and to provide avenues for future research. Moreover, validation of the prototype's practical suitability by the industrial experts can further bridge the gap between industry and academia.
% Note that performance constraints are usually dependent on the underlying blockchain-based framework and are not considered separately. 

\subsubsection{Blockchain Type}
% /////BC type :Structure and Access Mechanism
%Structure: Chain, DAG
%Access Mechanism: Private, Public, Consortium
% Control: Permissioned Permissionless 

The blockchain type can be categorized with respect to structure (chain and Directed Acyclic Graph (DAG)) and with respect of access mechanism (public, private, consortium). The blockchain type must be selected depending on the underlying application requirements. However, the current DT-based blockchain solutions mostly adopt mainstream sequential chain-structured public or private blockchains, such as Ethereum (ETH) or Hyperledger (HLF). For instance,~\cite{angrish2018case, hasan2020blockchain, putz58ethertwin} use ETH while~\cite{zhang2020manufacturing, leng2019manuchain} use HLF as the underlying framework for DTs. The main reason for such trend is due to the increasing number of implementations targeting different application domains while other Distributed Ledger Technologies (DLT) may not have reached the required level of maturity in terms of implementation and may not be mature enough for mass adoption yet.

\subsubsection{Scalability}
% Challenges
The selection of apropos blockchain is of significant importance to avoid scalability snag that may arise due to the futuristic surge based on the growing number of actors and activities in the Cyber-Physical Systems (CPS). The existing solutions of solving the scalability of blockchains include on-chain solutions (increasing block size, improved consensus strategies, sharding, and blockchain data structure) and off-chain solutions (off-chain channel and side-chain)~\cite{zhou2020solutions}.
%Existing works
Among the current blockchain-based DTs solutions, scalability issues are addressed by off-chain solutions are (as discussed in Subsection~\ref{sec:datastorage}).
Considering the data structure of blockchain, the DAG-structured blockchains (IOTA~\cite{popov2017iota}, NANO~\cite{lemahieu2018nano}, Byteball~\cite{churyumov2016byteball}), lightweight chain-structured blockchains (LSB~\cite{LSB}, PoET~\cite{PoET}), or alternative structures (Tree-Chain~\cite{Tree-Chain}), must be considered.
% ///////

\subsubsection{Smart Contracts} \label{subsec:smartcontract}

% What is SC?
A smart contract is an automatically executable code running on the blockchain to control digital assets by enforcing an agreement between untrustworthy parties~\cite{wang2019blockchain}. Smart contracts can be deployed to automate scenarios depending on the underlying application requirements. 
% Connecting Smart contracts with DTs:
%Existing works
When deployed into a DT environment, the smart contracts can be used to track data sharing mechanism~\cite{huang2020blockchain}, to represent twin-creation transactions~\cite{hasan2020blockchain}, to store authorization information for all involved parties~\cite{dietz2019distributed, putz58ethertwin}, or to automate event-based interaction among machines~\cite{angrish2018case}.   
% Other examples
Other examples that may involve smart contracts utilization in CPSs are triggering S\&S rules (as discussed in Subsection~\ref{sec:SS}), invoking PLC functions due to changing states of physical processes, modifying simulation setup parameters,  or auditing DTs by actively or retroactively monitoring smart contract transactions.
% ///////

\subsubsection{Consensus Mechanism}

% What is Consensus mechanism?
To keep the ledger state synchronized across the network while establishing a secure trusted network among untrusted nodes, consensus mechanism is performed by the participating nodes. Consensus algorithm involve solving a resource-intensive hard-to-solve and easy to verify puzzle. We do not describe the working of consensus algorithms here, as there are other works describing different consensus strategies of blockchain, for example~\cite{bodkhe2020survey,zhou2020solutions}.
% Existing works: % Reason for adopting x type of CM
Among the existing works on blockchain-based DTs, Proof of Work (PoW) is used by~\cite{huang2020blockchain, khan2020towards} without stating any convincing reason for this adoption.~\cite{angrish2018case} tested both PoW and Proof-of-Authority (PoA) to support the argument that the consensus mechanism must be chosen based on the underlying application. For example, PoW exhibits high computing power in comparison to PoA whereas PoW is more suitable for partially trusted environment and PoA is well-suited for permissioned environment. 
However,~\cite{zhang2020manufacturing, leng2019manuchain} use Cross Fault Tolerance (XFT) to achieve higher throughput and lower latency. Similarly,~\cite{putz58ethertwin} suggested using resource-efficient byzantine-fault tolerant consensus algorithm such as
Istanbul Byzantine Fault Tolerant (IBFT) as a part of future improvements to their prototype.

% ///////////////
\subsubsection{Data Storage} \label{sec:datastorage}
%Existing works
In terms of data storage, most of the existing works suggest using off-chain storage~\cite{angrish2018case, dietz2019distributed, hasan2020blockchain}, however, do not properly implement it. For instance, Hasan et al.~\cite{hasan2020blockchain} suggested using InterPlanetary File System (IPFS) storage but the details on how IPFS is used, are still missing. In comparison to other works, Putz et al.~\cite{putz58ethertwin} implemented off-chain storage based on swarm Distributed Hash Table (DHT). Following this approach, the off-chain data storage can meet the data volume and latency requirements where sensor feed is stored off-chain after registering sensor on-chain.

The idea of off-chain storage can help to solve the scalability issues in addition to reducing on-chain storage costs in the traditional blockchain. However, integrating off-chain storage solutions with blockchain-based DTs have several key challenges, e.g., there is no guarantee on the validity of the off-chain transactions, off-chain storage requires centralization, state consistency with on-chain storage in a real-time manner, and security issues~\cite{wang2019sok}.

\subsubsection{Traceability}
%Existing Solutions
Though blockchain ensures traceability through its append-only nature, it requires the selection of key parameters to track and trace the entire chain throughout the process. Some of the existing works (\cite{dietz2019distributed,zhang2020manufacturing,hasan2020blockchain}) discuss traceability but do not illustrate its working mechanism or actual implementation.
% Challenges/Solution
To enforce traceability in the industrial system, provenance data based on $\langle source, destination \rangle$ pair and other key identifiers can be constructed (as suggested in~\cite{suhail2019orchestrating}) while traversing through the process and can be stored on the blockchain to reconstruct the process chain on demand.

\subsubsection{Trustworthy Sources} \label{subsec:trustworthysources}
%Existing Solutions
While stringent security solutions are inherited from the blockchain, ensuring the~\emph{trustworthiness} of data-generating sources and data in transit is, by far, not considered by most existing works. 
% Challenges
Maliciously or mistakenly, data adulteration may occur due to various reasons such as device tampering, data forging, or maliciously participating entity thereby causing the Garbage In Garbage Out (GIGO) problem~\cite{suhail2019orchestrating}. 
Given that blockchain mechanisms do not guarantee the trustworthiness of data at the origin,~\cite{dedeoglu2019trust} proposed a blockchain-based layered trust architecture for IoT for improving end-to-end trust. In this proposed framework, at the data layer, the data trust module quantifies the confidence in observational data based on (i) evidence from neighboring data sources, (ii) reputation of data sources depending on their long-term behavior, and (iii) confidence level of the observation reported by data sources.
% Solutions
To address such issues in blockchain-based DTs, Suhail et al.~\cite{suhail2020trustworthy} suggested the following measures: 
\begin{itemize}
\item To circumvent device tampering, the sensor nodes are required to register as authorized devices through their public keys and digital certificates in the blockchain. Such measures enable the gateways to collect data from authorized devices only by first confirming their authenticity from the blockchain. Furthermore, firmware updates are also carried out through blockchain to avoid malicious code injection or malware.
\item To diagnose data forgery during data transition, potential solutions involve, i) reliance on provenance data to identify the faulty entity in the underlying chained processes and ii) leveraging twinned data to identify anomalies in the system that can be detected based on the digital-physical mapping. 
\item To identify a rogue participating entity through, i) issuing or revoking participation certificates by conducting a periodic on-site inspection of the units to ascertain that industrial processes are abiding by environment-friendly operations and ethical practices, and ii) deploying a  monetary punishment mechanism~\cite{ramachandran2017using} to reward honest entities while penalizing or revoking trader's participation.
\end{itemize}

% /////% /////% /////End of part II

\subsection{Design Solutions for Security}
In this subsection, we discuss the existing works pertaining to security-related design issues including encryption, access control, and quantum resistance. 

\subsubsection{Encryption}
% Existing work
Ensuring confidentiality through encryption is proposed by some of the works but is not described in detail or properly implemented. For example, Huang et al.~\cite{huang2020blockchain} suggested encrypting the product lifecycle
data; however, the implementation details of such encryption mechanisms are not provided which makes it hard to analyze. %does not provide implementation details. 
Similarly, Putz et al.~\cite{putz58ethertwin} encrypt data before uploading it to the DHT using AES-256 whereas permissions are enforced by encrypting a symmetric key with the corresponding entity's public key.
% Challenges
It is also an established fact that the security of de-facto cryptographic primitives is at risk of being broken by quantum computers~\cite{cheng2017securing}. Therefore, to address this critical issue, one set of solutions is to use quantum-resistant public-key cryptography (also called Post-Quantum Cryptography (PQC)) as drop-in-replacements~\cite{chen2017cryptography}. Based on mathematical problems other than factoring and discrete logarithms (used by the traditional cryptographic algorithms), such solutions are yet believed to be secure against quantum attacks. Another set of solutions is quantum cryptography, most notably, Quantum-Key Distribution
(QKD), which is a method for establishing symmetric
keys~\cite{mosca2018cybersecurity}. However, due to certain limitations (i.e., presence of quantum channel for sending bits), it is difficult to use such a solution as a plug-and-play replacement for the current security methods~\cite{mulholland2017day}.

\subsubsection{Access Control}
% Existing works
The current literature on blockchain-based DTs has a few authentication and authorization mechanisms as the concomitant of access control. 
However, we could not find their implementation details. % are mentioned by a few works but are not implemented. 
For instance, Dietz et al.~\cite{dietz2019distributed} proposed a stored role-permission mapping to grant or deny access based on Role-Based Access Control (RBAC). Similarly, Putz et al.~\cite{putz58ethertwin} implemented %and follow 
a hybrid access control model by combining RBAC and Attribute-Based Access Control (ABAC).
% Challenges
The access control capability provides authentication and authorization to enable data sharing among the involved parties and therefore must be considered carefully.

\subsubsection{Quantum Resilience}
% Challenge
% Quantum computers put blockchain security at risk by enabling wrongdoers to forge digital signatures and sabotage transactions~\cite{fedorov2018quantum}. 
With the arrival of quantum computers, wrongdoers are able to forge digital signatures and sabotage transactions thereby putting blockchain security at risk~\cite{fedorov2018quantum}.
To prepare for the quantum apocalypse, blockchain-enabled schemes that already support post-quantum techniques such as IOTA~\cite{popov2017iota} (using WOTS), Quantum Resistant Ledger (QRL)~\cite{QRL} (using XMSS), and Corda (using BPQS: a single-chain variant of XMSS) should be considered for long term security~\cite{suhail2020role}. Moreover, to further enhance security and enable blockchain to become faster and more efficient, requires quantum-linked features including quantum Internet, quantum-safe communication, and QKD for both communication and computational processing of blockchain data~\cite{fedorov2018quantum}.
% Existing work
Among the existing works, Khan et al.~\cite{khan2020towards} proposed a variant of blockchain called Twinchain which replaces Elliptic Curve Digital Signature (ECDSA) with quantum-safe Hash-Based Signature (HBS). However, the proposed quantum-resilient blockchain does not provide sufficient details about the chosen HBS. To this end, there are still outstanding challenges and research avenues that could be investigated by the research community.

\begin{figure}[ht!]
\centerline{\includegraphics[width=3.5in]{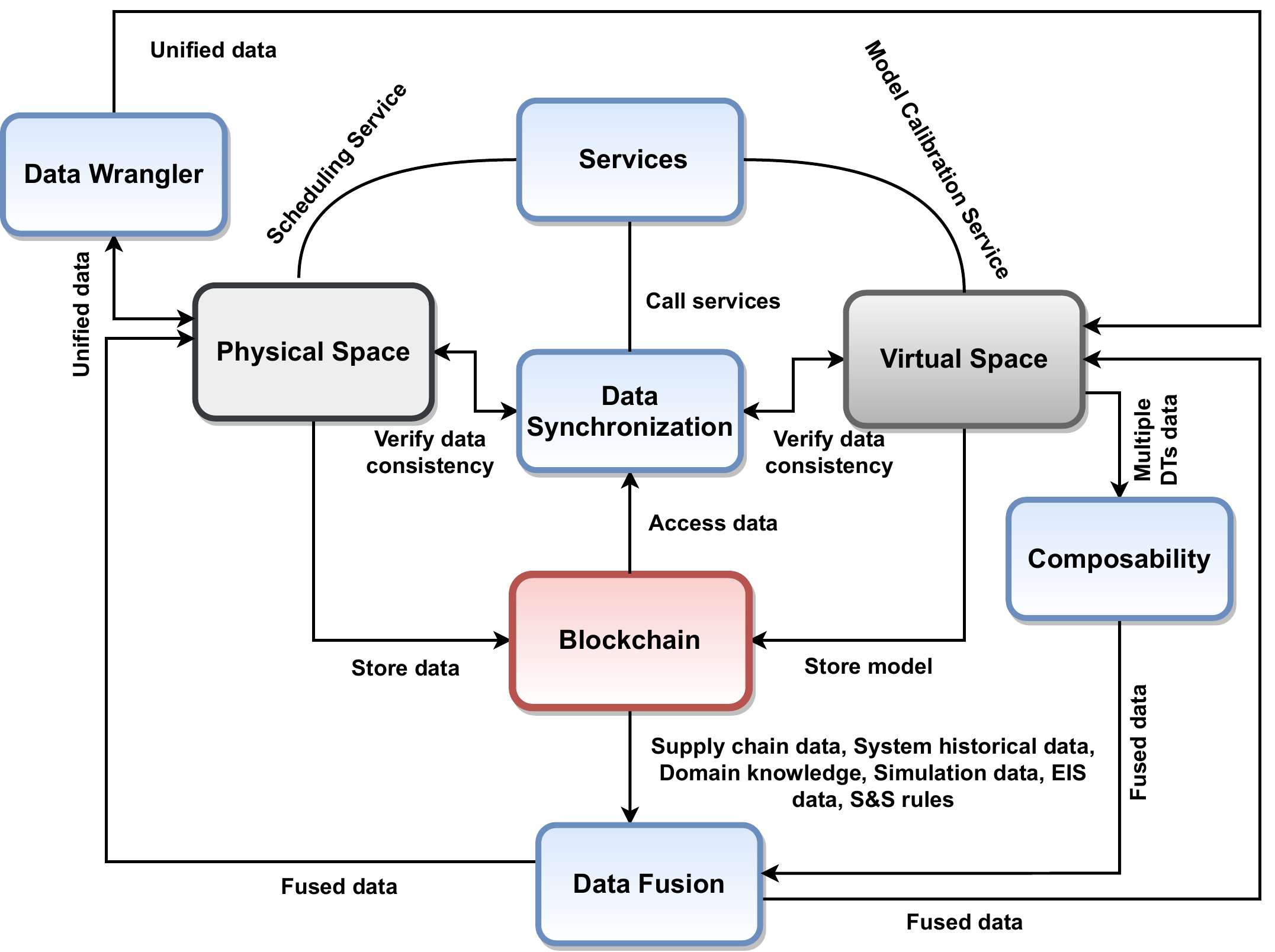}}
\caption{A trustworthy 7D DT model.}
\label{fig:7DModel}
\end{figure}

\begin{figure}[ht!]
\centerline{\includegraphics[width=4.5in]{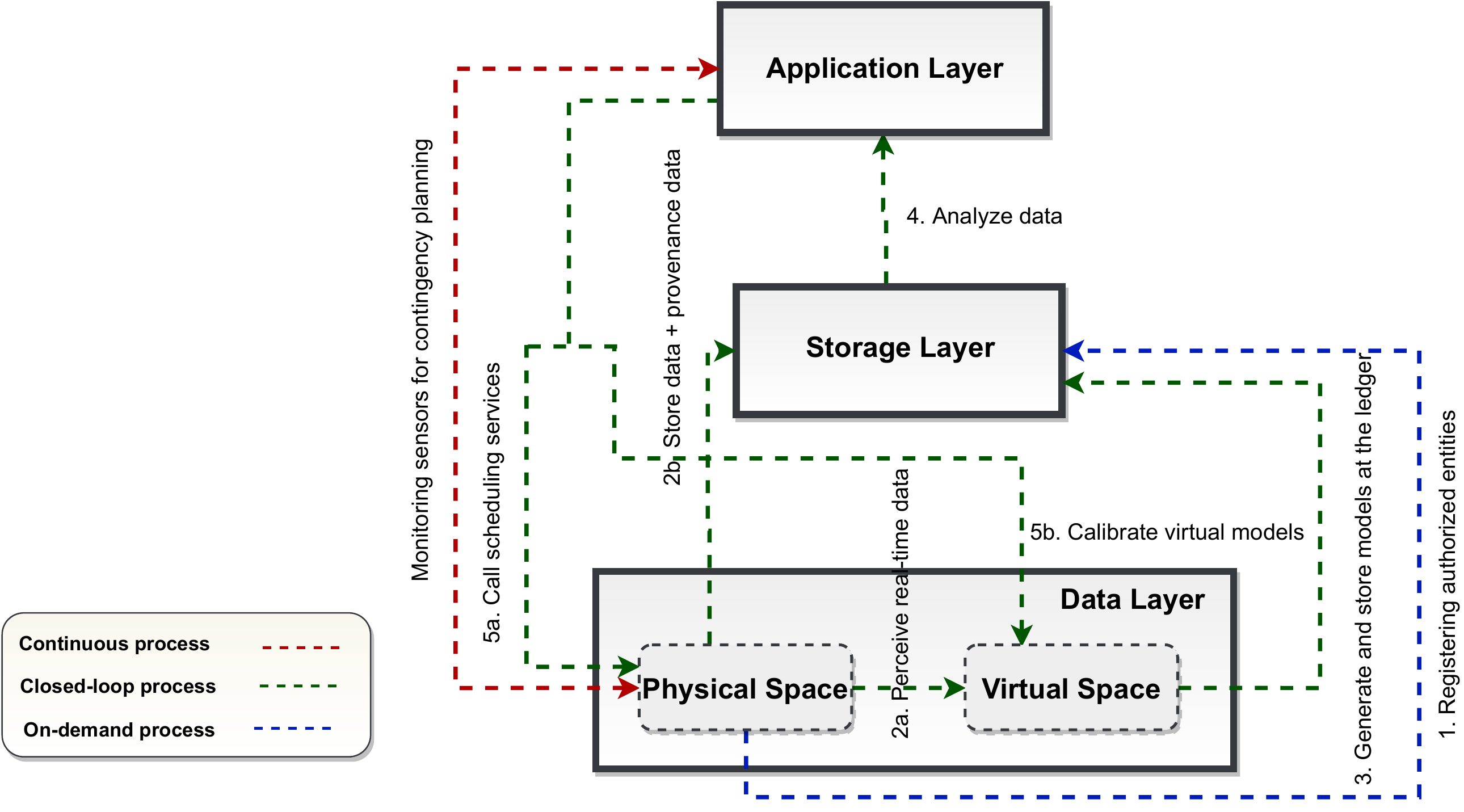}}
\caption{Blockchain-based digital twin: A framework. (Adapted from~\cite{suhail2020trustworthy})}
\label{fig:framework}
\end{figure}

In the following, we discuss a seven-dimension (7D) DT framework (Fig.~\ref{fig:7DModel}) based on the limitations of the existing works. We first provide a high-level layered structure of blockchain-based DTs (Fig.~\ref{fig:framework}). Then we further discuss the main and supporting components and their functionalities in connection with DTs-, blockchain-, and security-related design issues outlined in Table~\ref{tab:review_technicalwork}.

\subsection{Trustworthy Blockchain-based Digital Twins: A Unified Framework}
Curating and mining information from the collected data calls for an infrastructure that can manifest the trustworthiness of data-generating sources, data in transit, and data at rest (storage)~\cite{suhail2020trustworthy}. In this regard, Suhail et al.~\cite{suhail2020trustworthy} recently addressed the requirements of trustworthy data management, data security, and predictive maintenance through a provenance-enabled blockchain-based DT framework. The main components of the proposed framework are presented in Fig.~\ref{fig:framework}. At first, the participating entities such as sensors, machines, and humans register as authorized entities at the blockchain (step 1). Next, at the~\textit{data layer}, cyber-physical mapping is performed by monitoring, collecting, and processing designated parameters at the physical space that are mirrored at the virtual space (step 2a). The collected data along with provenance (i.e., a complete lineage of data along with a set of actions~\cite{ZAFAR201750}) is sent to the~\textit{storage layer} (step 2b). At the twinned system, models are generated based on collected data, domain knowledge, and system historical data and are stored at the storage layer (step 3). The~\textit{application layer} continuously analyzes data to detect manufacturing disturbances (step 4) and calls for the scheduling services (step 5a) or model calibration services (step 5b), thereby completing the agile feedback loop. Additionally, physical security countermeasures can be exercised through monitoring devices such as surveillance cameras, power monitoring, fire alarms, etc. Thus, blockchain ensures secure data management and DTs use provenance-enabled trustworthy data as input to predict events and their associated causes by means of data analysis~\cite{suhail2020trustworthy}. 

\begin{table}[!ht]
\renewcommand*{\arraystretch}{1.8}
\centering
\caption{Mapping DT properties with the proposed blockchain-based DTs model}
\label{tab:DTs_properties}
\begin{tabular}{ll}
\toprule
\multicolumn{1}{c}{\textbf{Properties}} & \multicolumn{1}{c}{\textbf{Blockchain-based DT model}} \\ 
\midrule
Clone counterpart & Virtual Space \\
Entanglement &  Data Synchronization  \\ 
Contextualization & Composability \\ 
Memorization &  Blockchain \\ 
Availability & Blockchain  \\ 
Data Integration and Interoperability &   Data Fusion, Data Wrangler  \\ 
Aggregation &  Composability \\ 
Secure Data Management and Accountability & Blockchain, Services \\
Input (Historic) Knowledge Base & Blockchain \\
\bottomrule
\end{tabular}
\end{table}

In the following, we discuss the 7D blockchain-based DT model which enhances the functionalities of the 5D DT model (shown in Fig.~\ref{fig:5Dmodel}) by introducing data integration and interoperability components (such as data fusion, data wrangler, and composability) and data synchronization. Also, it introduces blockchain (as an alternative to DT Data) to facilitate secure data management to all components. Table~\ref{tab:DTs_properties} presents a summary of the blockchain-based DT model by mapping them on the desirable properties of DTs. Furthermore, we map the high-level functionalities of the components in 7D blockchain-based DT model to the desirable characteristics outlined in Table~\ref{tab:review_technicalwork}. Finally, to illustrate the working of the proposed framework, we mapped the components of 7D DT model (Fig.~\ref{fig:7DModel}) to the production process of autonomous vehicles as a CPS use case. The proposed framework can be significantly beneficial for the safe (or fail-safe) operation of autonomous vehicles where safety and reliability are highly deemed.

\subsubsection{Design of Digital Twin}
In addition to the existing components of the 5D DT model (Fig.~\ref{fig:5Dmodel}), Suhail et al.'s framework~\cite{suhail2020trustworthy} introduces the following three components: (i) data wrangler, (ii) data synchronization, and (iii) data fusion. 
{\it Data wrangler} is responsible for cleaning erroneous (invalid or missing) data and converting different data formats into a unified format before inputting data to the physical and virtual spaces as mentioned in subsection~\ref{data_quality}.
For instance, during the vehicle production process, data from multiple physical sources such as sensors and actuators (attached to the robotic arms or vehicles structures) is gathered and is inputted to both the physical and virtual spaces.
{\it Data fusion} is used to combine data from multiple sources to generate consistent interpretation for a certain object. For instance, before initiating the production process of vehicles, the necessary details such as order information (e.g., material stock, production quantity, estimated cost), supply chain data (e.g., consignment information), simulation data (equipment historical data, prediction of equipment fault), thresholds and consistency checks (S\&S rules), etc. are already available in the storage system for usage. Furthermore, data wrangler and data fusion play a significant role in collecting and aggregating data after the vehicle production process. For example, to provide rich data about the car environment, the sensors attached to the connected vehicle (such as camera, GPS, infrared detectors, etc.) accumulate data during or after drive and then fed to the DTs for optimizing the operational control~\cite{suhail2021securing}. Such data can be used by threat intelligence to study the adversarial space and to update S\&S rules for stealthy attacks. To facilitate the creation of an abstract view of overall phenomena, DTs must support the correlation of different DTs associated with different physical sub-components. For instance, different physical process during the vehicle production process along with their twins must be combined. To do so, the framework uses {\it composability} that aggregates multiple DTs (i.e., DTA as discussed in Subsection~\ref{subsec:DTinstances}) to integrate and interoperate multiple DTs or replicas.

To register any instantaneous change (based on requirements of application) in the system status, DTs and their physical counterparts must be~\enquote{constantly connected}.
To do so, {\it data synchronization} enforces self-adaptation by performing a continuous digital-physical mapping between the pre-defined system performance parameters retrieved from the storage, and real-time sensor data obtained from the manufacturing unit, to verify data consistency. For instance, during manufacturing of vehicles, to minimize the makespan and production cost, DTs keep monitoring and analyzing the physical machine's current status based on constraints such as tool wear data, machine performance in the ongoing process, etc. To make DT reflecting the behavior and the status of the physical object, an effective and reliable communication suitable to the rate of changes are among other desirable properties to support entanglement between the physical space and the DT~\cite{minerva2021digital}. Depending on the data anomalies (defined in S\&S rules) or uncertain data inconsistencies, the service component calls the corresponding scheduling services (in the physical space) or model calibration service (in the virtual space) to carry out the necessary measures. Based on the above-mentioned phases, the lifecycle of DT evolves from the early phase (design and development) to the medium phase (operation and maintenance). Whereas the existence of DT may occur before the creation of the product and even after the end of product life. The existence of DT may occur before the creation of the product and it may continue to exist even after the end of product life. Usually, the dismissal of the object is followed by the dismantling of the DT.

\subsubsection{Design of Blockchain}
The storage layer provides secure distributed data storage through a lightweight, scalable, and quantum-immune blockchain. Data can be securely accessed to and from blockchain, for example, the application layer uses it for analysis and decision-making while DT uses it for generating and uploading updated virtual models. Also, the data synchronization can access the updated data directly from the ledger, to eliminate the qualms of untrustworthy data. Empowering DTs with blockchain solves the key issues, i.e., disparate data repositories, and untrustworthy data dissemination; however, to reason about the current {\it source} and {\it state} of a data object entails a complete lineage of processes chain. Therefore, to diagnose data forgery during data acquisition and transition and to support~\enquote{track and trace solutions}, Suhail et al.'s framework relies on provenance data and twinned data to identify the cause of deviations between the physical-digital space in addition to data assessment. 

Considering big data storage issues, we need to carefully consider what type of data needs to be stored on the blockchain. For instance, storing sensor data streams may not be an ideal situation due to data velocity and volume. The proposed framework relies on a data synchronization process that limits the frequent time-consuming access to the storage system, where the data flow of real-time sensor data and the less dynamic production and provenance data are explicitly separated. 

\subsubsection{Design of Security Features}
Currently, most of the blockchain-based solutions heavily rely on de-facto cryptographic primitives. However, blockchains that are not quantum-resistant could pose long-tail data risk by potentially enabling attackers with high-powered quantum computers to monopolize the network by forging digital signatures and sabotage transactions~\cite{fedorov2018quantum}. 
Therefore, to prepare for the quantum world, blockchain schemes (e.g.,~\cite{popov2017iota}) that support post-quantum cryptography must be adopted~\cite{suhail2020role}. Also, to allow multi-party use of DTs data, distributed data accessibility and auditability based on ownership, roles, and access levels must be enforced.

\section{Integrating Blockchain-based Digital Twins with AI: Way Forward} \label{DT_AI}

\begin{figure}[!ht]
 \centerline{\includegraphics[width=3.0in]{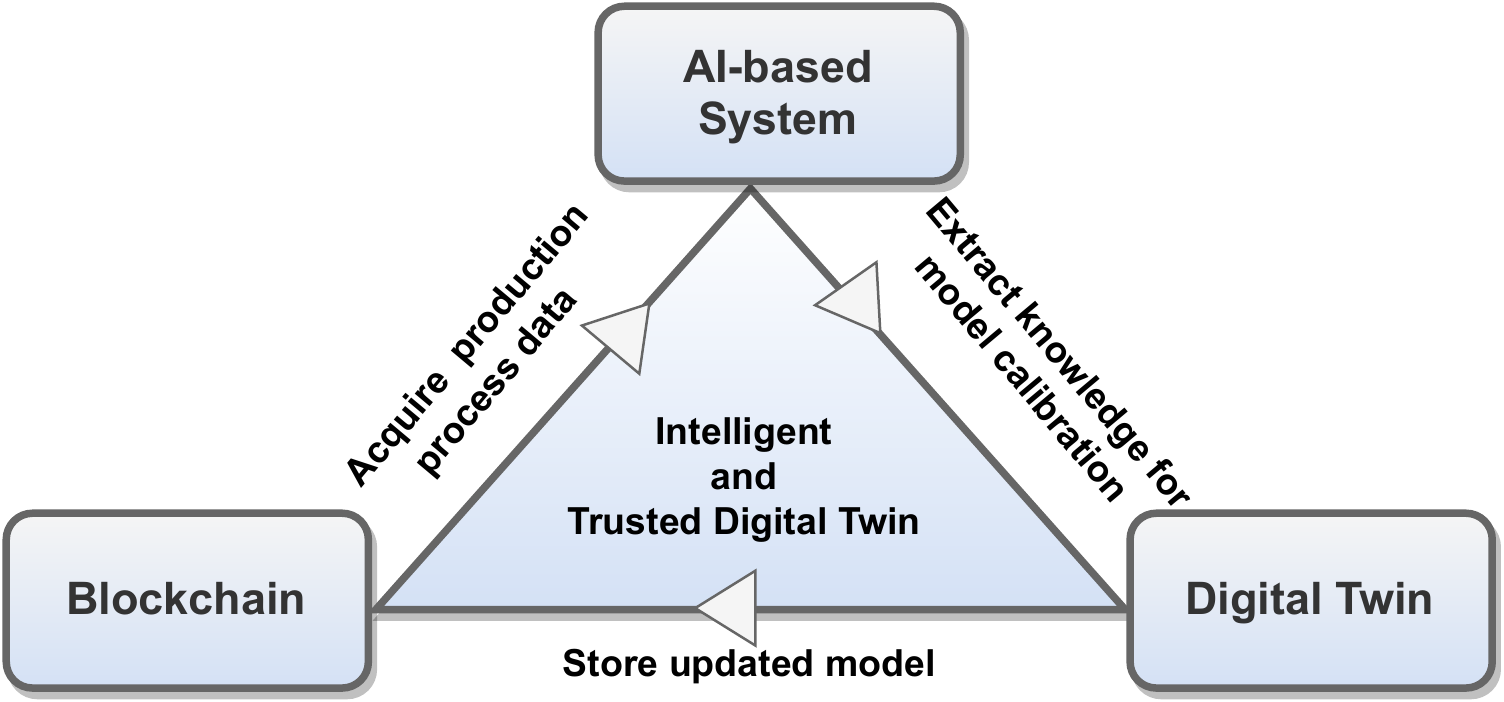}}
\caption{Integrating AI and blockchain for intelligent and trusted DTs.}
\label{fig:BC_DT_AI}
\end{figure}

Although DTs offer a variety of advantages, they may pose limitations. For instance, {\it how to analyze the huge volume of data for creating actionable insights in real-time~\cite{suhail2021securing}?} AI can address these limitations through the predictive capability of Machine Learning (ML) algorithms. For instance, blockchain and AI can provide intelligent and trusted DTs as shown in Fig.~\ref{fig:BC_DT_AI}. However, integrating AI for blockchain-based DT requires consideration of several aspects. In this section, we provide a way forward for AI-driven blockchain-based DTs. We start with the rationale for AI in DTs and blockchain. Then we look into the role of AI in blockchain-based DTs in a holistic way and the outstanding challenges that need to be addressed.

% % /////////////////

\subsection{How AI Will Set a Course for Blockchain-based Digital Twins}
In the following, we discuss some of the benefits brought by AI in blockchain-based DTs.

% Rearranged:
\subsubsection{AI-based anomaly detection solutions for DTs}
% Start: What is the PROBLEM?
The complex, heterogeneous, and interconnected nature of smart CPS is prone to errors and attacks due to the integration of multiple un-trusted but interdependent participating entities and disjoint data from a number of sources. 
% As the interdependence of the underlying environment increases, predicting the effects of apparently innocent actions or events could become more challenging. Whereas ignoring such small variations in the system behavior may collapse the whole system or may incur drastic effect on the long-term behavior of the system analogous to the~\enquote{butterfly effect} where a small change in a complex system might cause large effects on the overall system.
% Rewrite above lines:
For example, sequential interdependence of processes (such as assembly line) or reciprocal interdependence (such as vehicles passing back and forth between multiple units during manufacturing). Due to the increasing interdependence of the underlying environment, predicting the effects of apparently innocent actions or events could become more challenging. Whereas ignoring such small variations in the system behavior may collapse the whole system or may incur drastic effect on the long-term behavior of the system. 
% Example: drastic effect on long-term behavior of system
One such example within the ICS is Stuxnet where, without being detected, malicious code is able to intercept and modify the data sent to and from the PLC.

% What is the SOLUTION? AI (TI)
In this regard, implementing intelligence-driven solutions (such as data analytics and threat intelligence) provide autonomic security solutions to detect zero-day threats and can help to identify vulnerabilities, faulty models, threat actors, existing and potential attack vectors, thereby minimize the threat landscape~\cite{ibrahim2020challenges}. Additionally, threat intelligence covers the broader perspective of {\it `who'}, {\it `why'}, {\it `when'}, {\it `how'}, and the {\it `magnitude of damage'} about the data breaches and attacks. In literature, ML has proven to be useful in detecting security threats by analyzing security and log data. For instance, Castellani et al.~\cite{Castellani_2020} used weak supervised learning in DTs for anomaly detection in the industry settings. The training data is generated using DTs whereas labeled anomalous data is generated from the physical components in the industry. More precisely, the authors used Siamese AutoEncoders, especially tailored for weak supervised learning with synthetic data to detect anomalies in the real data.

% What is our strategy: Highlight AI with DTs and BC 
% Last: Conclusion of 4.2.1
With reference to the proposed model represented in the Fig.~\ref{fig:7DModel}, it is prudent to employ ML models in DTs data synchronization and model calibration services to ensure that stand-alone components have the ability to realize when it is being fed incorrect or malicious external data, detect the occurrence of an attack, and carry out a safe mission in the face of the attack~\cite{suhail2021securing}. Furthermore, the S\&S rules must be stored and retrieved from the ledger to ensure their reliability. For instance, a provenance-aware blockchain-based system can track and trace the accountable entity for adding or updating the S\&S rules~\cite{suhail2021securing}. Thus, based on the~\enquote{learn and prevent} mechanism, a blockchain-based DT governed by an intelligent AI-based system can detect the presence of malicious or mistaken disruptions in the system and can invoke the appropriate defence mechanisms automatically. Even if the damage is unavoidable, the AI-based solution, at the very least, isolates the attacked physical (sub) component, thereby keeping the rest of the infrastructure safe from further damage attack. 
 
% ////////////////////////////

\subsubsection{Improving Data Quality} \label{data_quality}
In addition to predictions, AI helps in improving the data quality by implementing intelligent automated methods to fill in the missing data and to clean the data (such as data denoising, data de-duplication). 
% Connecting with DTs 
For instance, AI-enabled data curation at the data wrangler method can help in achieving data completeness and data accuracy which ultimately improves the data analytics and optimization through the twinned system. To this end, Generative Adversarial Networks (GANs) can be used to improve the data quality by imputing missing data~\cite{Yoon2018}.
% Data augmentation 
Similarly, data augmentation can be exploited to expand the amount of relevant data by adding minor alterations to the existing data. 
Through algorithms such as neural network, Kalman filter, and bayesian inference, data collected from multiple heterogeneous sources can be fused to generate consistent and comprehensive interpretations of the object~\cite{tao2017digital}.
% Connecting with BC
While data quality can be improved through AI, data traceability can be enforced through blockchain-based solutions as discussed in Subsection~\ref{subsec:challenges}.

\subsubsection{AI-aided Predictive Maintenance in DTs}

% Connecting with DTs
Creating future scenarios in the absence of direct observation data is very challenging for DTs. In this regard, to predict anomaly detection or to perform risk assessment, the predictive capability of ML algorithms can play an important role. 
% Examples:
% For instance, supervised learning algorithms (such as LSTM) can be used for projections, predictions, or conducting sensitive analysis in a system through DTs. Other examples include unsupervised learning algorithms (such as k-means methods) and deep learning (such as reinforcement learning).
For instance, supervised learning algorithms (such as LSTM), unsupervised learning algorithms (such as 
% k-means methods
autoencoder), and deep learning (such as reinforcement learning) can be used for projections, predictions, or conducting sensitive analysis in a system through DTs. The appropriate algorithm can be chosen depending on the availability of labelled and unlabelled data in DTs. However, in such data-deprived situations, deep learning can be used.
% Connecting with BC
The blockchain can be used to record parameters and hyperparameters of ML and to log outcome in order to pinpoint trends in correct predictions.

\subsection{Challenges for AI in Blockchain-based Digital Twins} \label{subsec:challenges}
Despite the afore-mentioned benefits of AI in blockchain-based DT, there are still challenges that need to be addressed before the realization of AI-driven blockchain-based DTs. In essence, we focus mainly on the data-related challenges faced by AI in blockchain-based DT. Identification of these challenges is important to avoid pitfalls of AI in such environments.

\subsubsection{Need for Explainable AI} \label{subsec:XAI}

% Why XAI is important?
Despite the wide success of AI-based systems that are capable of perceiving, learning, predicting, and acting autonomously, there is a reluctance to adopt these systems in practice due to the black-box nature of most of the AI model. Decisions made by the resulting systems suffer from a loss of transparency and comprehensibility, particularly for the users who are not ML experts and hence can lead to a negative impact on the trustworthiness of the system~\cite{suhail2021securing}. Hesitation is even higher in safety-critical systems where explainability becomes crucial as slight dysfunction of an AI algorithm in DTs may lead to wrong decisions in the physical counterpart that could imply loss of life or economic disaster. Therefore, it is essential to remove the opaqueness of black-box-models by explaining their inner working. In this regard, explainable AI provides a set of techniques and methods to convert the black-box models to white-box approaches such that the overall process adopted by the algorithms and models to reach a decision can be explained and interpreted~\cite{hussain2021explainable}.
% Connecting AI with BC
However, democratizing the data require decentralized AI while assuring traceability and solidity of data and algorithms. In this regard, consolidation of AI and blockchain supports transparency for AI operations by tracking data flow and complex behaviours of AI-based systems to establish more understanding and confidence in the decisions made by the underlying systems~\cite{dinh2018ai}. The application of AI with federated learning algorithms is one such example that requires data traceability favored by blockchain to lurk privacy issues. 

% ///////
\subsubsection{Data Scarcity}

% Reason I: Data is scarce
Data serve as the lifeblood of AI models. To be able to learn the behavior and make accurate predictions, a sufficient (large) amount of data is necessary and oftentimes becomes the bottleneck for AI. Abstractly, there are two ways to generate data: (i) acquire data from the real environmental components (for instance sensors and actuators), and (ii) generate data synthetically (for instance, using GANs). 
% Reason II: Data is available but only accessible or owned by big companies
Another reason for data scarcity might be siloed data by tech-giants or reluctance to share data due to competition.
% How data scarcity problem in AI can affect DTs: Connecting AI with DTs
Since DTs consumes real-time and historical data to simulate or replicate the real-world environment, the use of AI in DT should take data scarcity into account. Uniform, consistent, healthy, and on-time data sources must be present to be fed to the AI models. 
% How BC can solve data scarcity problem? Connecting with AI and BC-based DTs
% Giving incentives to companies and users for sharing
One possible solution to solve data scarcity problem is to incentivize users for crowdsourcing and crowdselling data, for instance, through blockchain-based data marketplace, streamr~\footnote{https://streamr.network/}.
%Answer to Raja comment: how would crowdsourcing work in a digital twin industrial scenario?
The industry stakeholders may utilize the crowdsourced data during early phase (design and development) of DTs lifecycle. Similarly, the crowdsourced data may provide hypothetical conditions for simulation mode particularly when the current physical state of the system is not known. 
% federated learning
Federated learning can be used to solve the data scarcity problems by means of training local models at the user's devices with the user's private data.

\subsubsection{Model Errors and Bias Problem}

% What is the problem?
Model errors are closely coupled with the bias problem and are inherently related to the data inputted to AI. If the AI model is trained on the imbalanced or poor quality data, then the outcome of the learning can be biased thereby causing prediction errors and compromising the efficacy of AI. 
% How this problem can affect DTs?
Since the data from both the physical and its twinned environment is used for learning, it is important to consider the source of data, the environment in which it is collected, the context of data collection, and the way the data was collected to avoid biasness in data and eventually model errors.
% Connecting with blockchain
To enforce data quality, employing blockchain can solve the bias in data, and model error problems in two ways: (i) incentivizing the participants to contribute good data~\cite{justin2019decentralized}, and (ii) 
gathering data from multiple siloed sources 
% rather individual sources
through blockchain-based federated learning~\cite{ur2020towards}.
It is important to note that incentivisation of contributing data will depend on the specific blockchain type used and the built in incentives to contribute data. For data bias through sharing data from multiple sources, this in itself does not solve the problem. Data fusion methods that incorporate trust are needed on top of blockchain to overcome the data bias problem. 
Additionally, collective decision-making processes can be carried out through decentralized infrastructures and consensus protocols. Thus, human users can re-calibrate their trust in the system by tracing back the decision process and making justifications that are essential to identify which entity (human or machine) is faulty.

%///////////////

% //////////////////////////////////////////

\section{Research and Deployment Challenges} \label{challenges}
\begin{figure}[ht!]
\centerline{\includegraphics[width=5.0in]{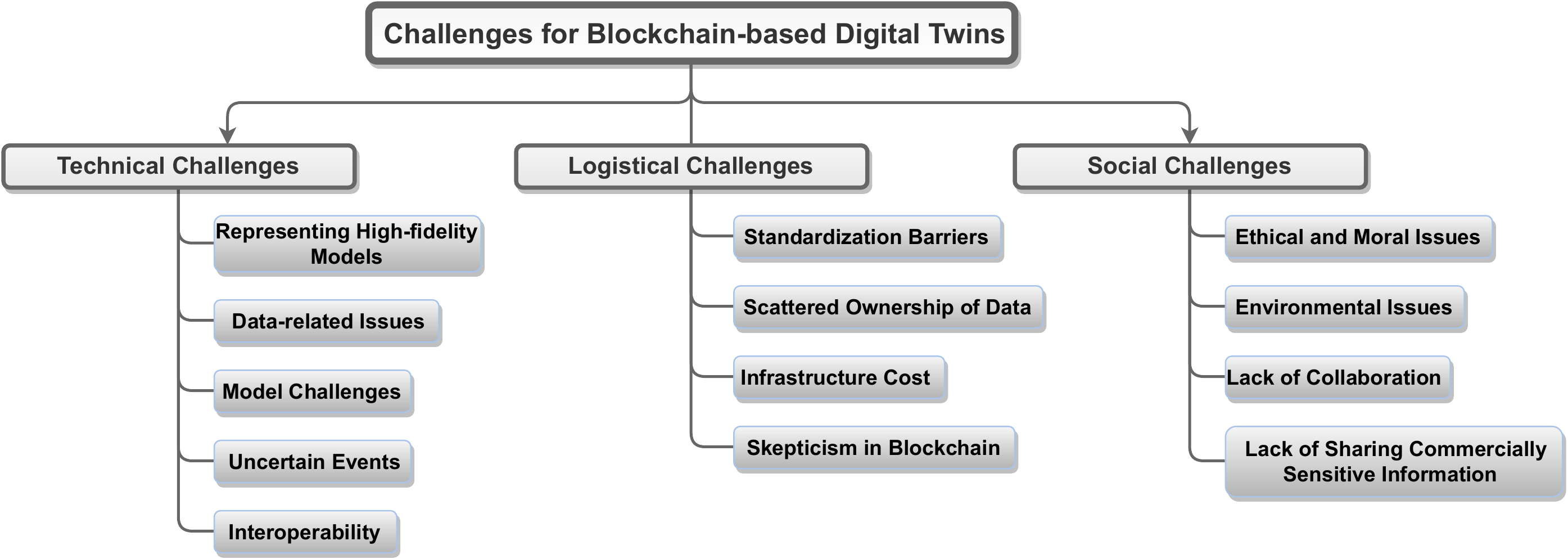}}
\caption{Current and future challenges for blockchain-based digital twins in the industrial domain.}
\label{fig:challenges}
\end{figure}

In the quest to enforce secure data management in the Industry 4.0 era, the following technical, logistical, and social challenges of blockchain-based DTs schemes require further investigation. In this section, we highlight the open-ended challenges that hinder the successful implementation of blockchain-supported DTs in the industry. These challenges open new avenues for future research in these directions. Fig.~\ref{fig:challenges} presents the detailed taxonomy of the current and future research and deployment challenges for blockchain-driven DTs. We summarize the challenges along with causes and possible solutions in Table~\ref{tab:technical_non-technical}.

\subsection{Technical Issues}
Here we discuss technical challenges with reference to blockchains, DTs, and their integration.
\subsubsection{Representation of Digital Twins}
% 1. High-fidelity
\enquote{High-fidelity} model is one of the critical challenges for a virtual model that represents the real-world situation. For instance, in a production unit, the quintessential DT for a robotic arm should mirror an exact image of all the behavior and functionalities of the physical component, synchronized in real-time throughout its lifecycle~\cite{suhail2020trustworthy}. However, in industry, uncertain scenarios transpire due to the dynamism and complexity of underlying (sub) systems, machinery malfunctioning, human errors, hidden vulnerabilities, erroneous data, knowledge gaps, and variability, fuzziness, and uncertainty of physical asset. Moreover, while maintaining backward compatibility, a corresponding evolution of the models is required as the physical object evolves over time~\cite{rasheed2020digital}. Therefore, to enable DTs to unearth the unknown unknowns, a pragmatic approach must be adopted by integrating cognitive DT solutions~\cite{ali2021cognitive}. 

% 2. Replication (we can talk about simulation or data analytics too)
Considering representativeness and contextualization in DTs, feature simplifications or generalizations are possible, provided that they do not affect the behavior in the specific context under a specific environment~\cite{minerva2021digital}. In this regard, replication modes are used for tracing physical objects and reproducing activities by emulating their behavior~\cite{dietz2020unleashing}. Initiating replicas in the framework or in the application space poses issues. For example, depending on the time period defined by the application, all replicas of a physical object should have the same status as that of the physical object. Otherwise, it may lead to the problem of synchronization and consistency of information between the physical object and its different replicas. To solve such issues, one possible solution is to integrate transaction management functions through blockchain. Other issues that may arise with replication are temporal dependencies and the need for input knowledge (such as events and activities) in advance to produce the same stimuli. Also, it requires experts to solve complex and intertwined problems.

% 3. interface issues
To emphasize socio-technical design implications, Barricelli et al.~\cite{barricelli2019survey} focused on Human Work Interaction Design (HWID) and End-User Development (EUD) to bridge the communication gaps among domain experts, designers, and stakeholders. In this regard, DTs must be equipped with well-designed and accessible interfaces such that it appears indistinguishable from the physical asset and allows end-user (domain experts or non-informatic experts) to intuitively create, modify, test, and operate digital artifacts.

\subsubsection{Data-related Challenges} 
Data challenges mainly emerge due to the following four factors: i) types of data to collect, ii) rate of data collection, iii) data storage, and iv) merging of disparate data types. The first issue requires a trade-off between excessive data volumes and limited data volumes, i.e., too few data may result in inaccurate predictions, and too many data may result in mired down in details or may pose decision paralysis due to information overload~\cite{suhail2020trustworthy}. The second issue involves the frequency at which data samples are collected. For instance, recording vibrations from a turbine gearbox every minute would miss shorter glitches but sampling every second could result in too much data thereby causing transmission bottlenecks~\cite{tao2019make}. Furthermore, the duration for which data is stored needs to be determined as high volumes of rapid measurements lead to high storage cost but at the same time, long-term data are required for modeling~\cite{kusiak2017smart}. Therefore, depending on the system requirements, it is important to define the spatio-temporal granularity of the data. The third issue corresponds to storing~\textit{big data} on the blockchain that requires careful consideration of factors such as high cost associated with high volumes of rapid measurements, block size, latency, etc. Alternative storage mechanisms and solutions include off-chain storage, storing only data hash to reduce storage costs, or removable blockchain designs~\cite{dorri2019mof}. Despite these solutions, the big data storage problem considering the volume and velocity of data are still open issues. The fourth issue involves disparate data types that are hard to merge, for instance, vibrations, temperatures, videos, or images all have different data types, formats, structure, and scale. Under such scenarios, missing or erroneous data can distort results and obscure faults, timings can get out of step during data sampling at different rates, and averaging fine data may lead to loss of details~\cite{tao2019make}. In a nutshell, to embed viable and size-efficient data into the ecosystem, the data-related challenges including volume, variety, veracity, generation rate, spatio-temporal resolution of sensor data, and efficient processing, archival, and retrieval require further investigation.

\subsection{Logistical Issues}
In the following, we outline some logistical challenges faced by the adoption of blockchain-based DTs in the industrial domain.

\subsubsection{Expenditures on Infrastructure} 
The adoption of emerging technological trends, such as IoT, High-performance computing (HPC), blockchain, DTs, AI, etc. aids in reducing operational costs, reinforce productivity, provide predictive and preventive maintenance of industrial machines, and attain higher operational efficiency. However, owing to such innovations, companies have to spend a considerable amount of capital on the day-to-day management of networked devices (maybe geographically dispersed). Such practices require the balancing of the cost-benefit analysis. Otherwise, it could drain operational resources for managing industry assets and business processes. Despite the significance of DTs in reducing risks, increasing safety, and improving efficiency, many enterprises realize that the connectivity, computing, bandwidth, and storage required to process massive volumes of data involved in creating DTs are cost-prohibitive~\cite{suhail2020trustworthy}. Likewise, for blockchain technology, perceived risks associated with the immature technology, high costs of initial implementation, and the likelihood of disrupting existing practices may pose significant challenges to businesses. Another addition to the Capital EXpenditures (CapEXs) and Operational EXpenditures (OpEXs) involves the issue to train security specialists and institutions with the state-of-the-art applied tools and knowledge of the data used to establish the model. Otherwise, improper or missing influencing factors can lead to fallacious results on which the critical security decisions and actions are based~\cite{dietz2020unleashing}. In such situations, shifting to the new paradigm is more beneficial for medium-to-large scale industries in contrast to small-scale industries. However, sooner or later, small-scale industries have to adopt the current technologies as they may scale in the future or ultimately due to technological obsolescence.

\subsubsection{Standardization Barriers and Efforts} 
% DTs
Due to the lack of applicable laws, the absence of unifying data and model standards, and integration problems into business processes,  an internationally operational framework and infrastructure for DTs is still a long way away. Typically incompatibilities exist throughout the entire development process and life cycle. For instance, scattered ownership of data involves different data interfaces, proprietary file formats, object definitions, encoding, and sharing of knowledge strewn throughout the product lifecycle. Moreover, it includes a multitude of vendors, contractors, and customers of all sizes and specializations that do not necessarily use the same software/tools, etc. Therefore, the impetus to create DT standard requires improved interoperability where each step of the workflow ingests artifacts from the previous stage and output artifacts that can be consumed by the following phase~\cite{piroumian2021digital}. With firm guidelines for cross-disciplinary standardization, a universal non-proprietary design and development platform for DTs should be developed with user-friendly and API-based interfaces on which all models can run without data monetization or fixed vendor partnerships.

The common denominator among blockchain platforms is their reliance on a distributed and decentralized peer-to-peer network. They nevertheless vary substantially in terms of the underlying data structure, scalability solutions, fault tolerance, degree of decentralization, and consensus approaches. The disadvantages due to non-interoperable blockchain implementations are manifold, for example, information restrictions, highly fragmented industrial ecosystems, stand-alone and disconnected networks, to name a few. In essence, allowing seamless integration of data by enforcing open data structure and processing harmonization through common linked data language and multiple agile, scalable and mergeable blockchain ledgers for logging different data types can minimize obstacles in the widespread adoption of blockchain. The Enterprise Ethereum Alliance (EEA), the International Organization for Standardization (ISO), the standardization sector of the International Telecommunications Union (ITU-T), the World Wide Web Consortium (W3C), and the Society for Worldwide Interbank Financial Telecommunication (SWIFT) carried out notable efforts in this regard.

\subsubsection{Skepticism in Blockchain Technology}
Many current research works are raising concerns about the practical adoption of blockchain technology in the industry. Some of their concerns include how to ensure the reliability of data from the participating entities and/or sensors~\cite{wust2018you} considering the connection between the physical and digital world, latency issues with the increasing number of nodes in the network~\cite{occam,prewett2019blockchain}, lack of information leading to the existence of gray markets~\cite{babich2019blockchain}, security concerns due to quantum computing~\cite{fedorov2018quantum}, lack of privacy and GIGO problem~\cite{babich2019distributed}, decision paralysis due to information overload, high energy consumption~\cite{zhao2019blockchain}, throughput and latency issues~\cite{lezoche2020agri}, lack of standardization and shifting to new infrastructure from legacy systems~\cite{sternberg2020struggle}, etc.~\cite{suhail2019orchestrating}. Paradoxically, blockchain technology supports the full gamut of industrial applications while struggling to address such challenges. 

Additionally, problems stemming from the blockchain~\enquote{black-box} nature force consumers to trust the integrity and fairness of the processes without understanding the technical underpinnings. Due to this reason, companies are hesitant to make use blockchain technology in DTs. Innovation hubs and maker spaces organize webinars, workshops, tutorials, and certifications that can enable the teams to learn blockchain technology and explore its use-cases in their respective areas, for example, learning platform provided by IBM~\cite{IBM_blockchain}.

\subsection{Social Issues}
There are several risks involved with the proliferation of the blockchain and DTs on an individual level which eventually summed up as issues in their integration. In the following, we discuss the social challenges faced by the integration of blockchain-based DTs. 

\subsubsection{Ethical and Legal Issues} 
Due to industry rivalry, companies often do not want to share commercially sensitive information with their competitors. Moreover, commercial pressures dissuade companies from sharing models, smaller firms lose out, resulting in monopolies or oligopolies. For instance, giant companies such as Siemens, GE, IBM, etc. own most of the DTs. Additionally, on one hand, due to a lack of common space between academia and industry, data scientists, cybersecurity experts, engineering, and business strategists cannot communicate and share knowledge and software. On the other hand, researchers and practitioners patent or even hoard knowledge, ensuing limited access to the data. 
Such problems call for a platform managed by government funding agencies, consortia of private corporations, or by a coalition of universities to reconcile issues about data governance, data ownership, and openness~\cite{tao2019make}. In this regard, a close-knit team of specialists spanning disciplines should collaborate, build precise DTs, and share data and models on a public database for widespread usage. Currently, such examples include openVertebrate (share data and models of vertebrate anatomies), Siemens NX software (package design, simulation, and manufacturing tools), LlamaZOO (develop interactive 3D visualization, digital twinning, and training solutions for Industry 4.0 through VR/AR), and IBM Model-Based Systems Engineering (streamline product design and development). Some of the open challenges such as how to resolve the ownership rights in scenarios related to DTs of retired or former human workers, or regulations for using healthcare data with consent and anonymously, still need to be addressed. Also, these challenges motivate the design of blockchains for DTs that protect sensitive information while providing sufficient transparency. 

\subsubsection{Environmental Impact} \label{subsec:EnvironmentalImpact}
Inefficient by design, blockchain consensus protocols are considered the Achilles’ heel in terms of energy constraints. The high cost incurred on the consensus phase is due to massive electricity consumption in performing compute-intensive mining phase (running cryptographic algorithms); draws a worrying figure for energy footprints. For example, with an estimated energy expenditure of almost 10 TWh, Ethereum matches the total energy consumption of some nations~\cite{digiconomist}. To curtail such an impact on the environment, possible solutions involve switching from energy-hungry to less power-consuming consensus mechanisms, such as TreeChain~\cite{dorri21} or deploying blockchains featuring miner-free solutions, such as IOTA~\cite{popov2017iota, Strugar2018}. Moreover, the AI-supported energy planning of blockchain-based systems could be optimized in how to allocate and maximize resources.

% % ///////
\begin{longtable}{b{4.0cm}m{5.5cm}m{5.5cm}}
\caption{Current and future research and deployment challenges in blockchain-based DTs.} \label{tab:technical_non-technical} \\
\toprule
\textbf{Class} & \textbf{Key challenges}  & \textbf{Possible solutions}     \\
\midrule
\textbf{T1:} Representing DTs & \begin{itemize} \item Accurate representation of physical object
\item Contextualization in DTs through replications 
\item Design implications
\end{itemize}
& 
\begin{itemize} \item Integrate cognitive DT solutions
\item Integrate transaction management functions
\item Focus on HWID and EUD
\end{itemize}
\\
\hline
\textbf{T2:} Data-related issues  & \begin{itemize} \item Types of data to collect 
\item Rate of data collection
\item Data storage
\item Merging disparate data types
\end{itemize}
& 
\begin{itemize} 
\item Consider data granularity
\item Off-chain storage, removable blockchain designs
\end{itemize}
\\
\hline
\textbf{L1:} Expenditure on infrastructure &
\begin{itemize} \item High cost on managing infrastructures 
\item Training personnel cost
\end{itemize}
&
\begin{itemize} 
\item Plan expenditure on software and hardware costs
\end{itemize}
 \\
\hline
\textbf{L2:} Standardization barriers &
\begin{itemize} 
\item Absence of unifying data and models
\item Scattered ownership of data
\item Non-interoperable blockchains
\end{itemize}
&\begin{itemize}
\item Developing universal platform for DTs
\item Integration of data among different blockchains
\end{itemize}
 \\
 \hline
\textbf{L3:} Skepticism in blockchain technology &
\begin{itemize} 
\item Black-box 
\end{itemize}
&\begin{itemize}
\item Supporting learning platforms 
\end{itemize}
 \\
\hline
\textbf{S1:} Ethical and moral issues & 
\begin{itemize}
    \item Lack of commercially sensitive information 
    \item Companies owning DTs may monopolize global market
    \item Limited access to data and models
    \end{itemize}  
    & \begin{itemize}
    \item Collaborating and encouraging widespread knowledge of data and models
    \end{itemize}
     \\
\hline
\textbf{S2:} Environmental issues & 
\begin{itemize}
    \item Energy consumption due to compute-intensive mining phase
\end{itemize} 
& 
\begin{itemize}
    \item
    Use of efficient consensus mechanism 
     \end{itemize}
    \\
\hline
\multicolumn{3}{c}{{\textbf T}: {\it Technical challenges}  \hspace{10pt}  {\textbf L}: {\it Logistical challenges} \hspace{10pt} {\textbf S}: {\it Social challenges}} \\ 
%  \hline
\bottomrule
\end{longtable}
% \end{landscape}
% ////////////////////

% \section{Concluding Remarks}\label{conclusion}
\section{Discussion and Conclusion} \label{conclusion} Due to the involvement of multiple entangled entities and ongoing intricate industrial processes, industries have to consider several challenging factors for maximizing their revenues, such as cost management, risk prognosis, and improving efficiency. Therefore, to maximize industrial efficiency, one of the promising solutions is to create a digital twin that mirrors every facet of the underlying system or process in order to analyze, predict, and optimize all operations before its real-world implementation. Driven by data collected and accessed from heterogeneous data sources and data silos owned by multiple participating entities calls for a secure distributed infrastructure. In this regard, leveraging blockchain allows industries to manage data on a distributed ledger that securely records product lifecycle data management events. Thus, DTs improve system maintenance by focusing more on preventing problems rather than on solving problems while blockchain provides the creation of secure DT frameworks. 

In this survey, we have conducted a detailed and comprehensive review of the design and implementation issues of the current state-of-the-art blockchain-based DTs solutions in the industry. The key takeaways from this survey are as follows:
\begin{itemize}
    % Ans to RQs:
    \item The integrating blockchain in DTs allows product lifecycle stakeholders to manage data on a shared distributed ledger to solve data trust, integrity, and security challenges.
    \item The comparative analysis of existing blockchain-based DTs identified the research gaps in the design solutions related to DTs, blockchain, and security. Based on the comparison shortcomings in the existing literature, a trustworthy 7D blockchain-based DT framework (Fig.~\ref{fig:7DModel}) has been proposed. 
    \item The integration of AI in blockchain-based DTs provide intelligent twins to support predictive maintenance and threat intelligence.
    \item Finally, the research community needs to investigate the research and deployment challenges mentioned in Table~\ref{tab:technical_non-technical} for the realization of the successful implementation of blockchain-based DTs.
\end{itemize}

We conjecture that this survey provides close insights to researchers to overcome the challenges and pave the way for the standardization of blockchain-based DTs in industrial applications.

\section*{Acknowledgments}
We thank the anonymous reviewers for their insightful comments and suggestions.

% https://www.acm.org/publications/authors/bibtex-formatting
% \bibliographystyle{ACM-Reference-Format}
% \bibliography{acmart}
%%% -*-BibTeX-*-
%%% Do NOT edit. File created by BibTeX with style
%%% ACM-Reference-Format-Journals [18-Jan-2012].

\end{document}